\documentclass[prd, aps, nofootinbib,preprintnumbers, showpacs,superscriptaddress,twocolumn]{revtex4}
\usepackage{latexsym}
\usepackage{amssymb}
\usepackage{amsfonts}
\usepackage{amsmath}
\usepackage[dvips]{graphicx}\usepackage{color}

\usepackage{ulem}

\renewcommand{\emph}[1]{{\it #1}}

\newcommand{\be}{\begin{equation}}  
\newcommand{\ee}{\end{equation}}
\newcommand{\bea}{\begin{eqnarray}}           
\newcommand{\eea}{\end{eqnarray}} 
\newcommand{\beqn}{\begin{eqnarray*}}
\newcommand{\eeqn}{\end{eqnarray*}}
\newcommand{\ba}{\begin{align}}
\newcommand{\ea}{\end{align}}

\def\de{\partial}
\def\lm{{\ell m}}
\def\g{{\gamma}}

\def\l{{\ell }}
\def\r{{\hat{r}}}

\def\u{{\bar{u}}}

\def\E{{\cal E}}
\def\B{{\cal B}}

\def\O{{\cal O}}
\def\X{{\bf X}}

\def\p4{{\psi_4}} 

\begin{document}


\title{Relativistic tidal properties of neutron stars}

\author{Thibault \surname{Damour}}
\affiliation{Institut des Hautes Etudes Scientifiques, 91440 Bures-sur-Yvette, France}
\affiliation{ICRANet, 65122 Pescara, Italy}

\author{Alessandro \surname{Nagar}}
\affiliation{Institut des Hautes Etudes Scientifiques, 91440 Bures-sur-Yvette, France}
\affiliation{ICRANet, 65122 Pescara, Italy}

\begin{abstract}
We study the various linear responses of neutron stars to external
relativistic tidal fields. We focus on three different tidal responses,
associated to three different tidal coefficients: (i) a gravito-electric-type
coefficient $G\mu_\ell=[\text{length}]^{2\ell+1}$ measuring the
$\ell^{\text{th}}$-order mass multipolar moment $GM_{a_1\dots a_\ell}$ induced
in a star by an external $\ell^{\text{th}}$-order gravito-electric tidal
field $G_{a_1\dots a_\ell}$; (ii) a gravito-magnetic-type coefficient
$G\sigma_\ell=[\text{length}]^{2\ell+1}$ measuring the $\ell^{\text{th}}$
spin multipole moment $G S_{a_1\dots a_\ell}$ induced in a star by an
external $\ell^{\text{th}}$-order gravito-magnetic tidal field
$H_{a_1\dots a_\ell}$; and (iii) a dimensionless ``shape'' Love number
$h_\ell$ measuring the distortion of the shape of the surface of a star by
an external $\ell^{\text{th}}$-order gravito-electric tidal field. All the
dimensionless tidal coefficients $G\mu_\l/R^{2\l+1}$,
$G\sigma_\l/R^{2\l+1}$ and $h_\l$ (where $R$ is the radius of the star)
are found to have a strong sensitivity to the value of the star's
``compactness'' $c\equiv GM/(c_0^2 R)$ (where we indicate by $c_0$ the speed
of light). In particular, $G\mu_\l/R^{2\l+1}\sim k_\l$ is found to strongly
decrease, as $c$ increases, down to a zero value as $c$ is formally extended to
the ``black-hole limit'' (BH) $c^{\rm BH}=1/2$. The shape Love number $h_\l$ is
also found to significantly decrease as $c$ increases, though it does {\it
  not} vanish in the formal limit $c\to c^{\rm BH}$, but is rather found to agree
with the recently determined shape Love numbers of black holes. The formal
vanishing of $\mu_\l$ and $\sigma_\l$ as $c\to c^{\rm BH}$ is a consequence 
of the no-hair properties of black holes. This vanishing suggests, but 
in no way proves, that the effective action describing the gravitational 
interactions of black holes may not need to be augmented by nonminimal 
worldline couplings.
\end{abstract}

\date{\today}

\pacs{
    04.25.Nx,   
    04.40.Dg,  
    95.30.Sf,  
  }

\maketitle

\section{Motivation and introduction}
\label{sec:intro}

Coalescing binary neutron stars are one of the most important (and most
secure) targets of the currently operating network of ground-based detectors
of gravitational-waves. A key scientific goal of the detection of the
gravitational-wave signal emitted by coalescing binary neutron stars is to
acquire some knowledge on the equation of state (EOS) of neutron-star matter.
Recent breakthroughs in numerical relativity have given example of the
sensitivity of the gravitational-wave signal to the EOS of the neutron 
stars~\cite{Baiotti:2008ra,Baiotti:2009gk,Kiuchi:2009jt,Read:2009yp}. However, this sensitivity 
is qualitatively striking only 
during and after the merger of the two neutron stars, i.e. for gravitational
wave frequencies $f_{\rm GW}\gtrsim 1000$~Hz, which are outside the most
sensitive band of interferometric detectors. It is therefore important to
study to what extent the gravitational-wave signal emitted within the most
sensitive band of interferometric detectors (around $f_{\rm GW}\sim 150$~Hz)
is {\it quantitatively} sensitive to the EOS of neutron stars. In such a 
regime, the two neutron stars are relatively far apart, and the problem can be
subdivided into three separate issues, namely:\\
(i) to study the response of each neutron star to the tidal field generated
by its companion;\\
(ii) to incorporate the corresponding tidal effects within a theoretical
framework able to describe the gravitational-wave signal emitted by
inspiralling compact binaries; and\\
(iii) to assess the measurability of the tidal effects within the signal seen by
interferometric detectors.

A first attack on these three issues has been recently undertaken by Flanagan
and Hinderer~\cite{Flanagan:2007ix,Hinderer:2007mb}.
[See also~\cite{Read:2009yp} for an attempt at addressing the third issue.]
Our aim in this work,
and in subsequent ones, is to improve the treatment of 
Refs.~\cite{Flanagan:2007ix,Hinderer:2007mb} on several accounts. The present
work will focus on the first issue, (i), above, namely the study of the tidal
response of a neutron star. Our treatment will complete the 
results of~\cite{Hinderer:2007mb} in several directions. First, we shall study
not only the usually considered ``electric-type'', ``tidal'', ``quadrupolar''
Love number $G\mu_2 = \dfrac{2}{3} k_2 R^5$, but also several of the other
tidal coefficients of a self-gravitating body. This includes not only the
higher multipolar analogues $G\mu_\ell \propto k_\ell R^{2\ell+1}$ of $\mu_2$,
but their ``magnetic-type'' analogues $G\sigma_\ell$ (
first introduced in~\cite{Damour:1991yw}), as well as their (electric) 
``shape-type'' kin $h_{\ell}$.
Second, we shall study in detail the strong sensitivity of these tidal
coefficients to the {\it compactness} parameter\footnote{To avoid
confusion with the compactness, we sometimes denote the velocity of 
light as $c_0$.} $c\equiv GM/c_0^2 R$ of the neutron star.
Note, indeed, that the published version of Ref.~\cite{Hinderer:2007mb} 
was marred by errors which invalidate the conclusions drawn there that
$k_2$ has only a mild dependence on the compactness $c$
(see e.g. Eq.~(27) or Fig.~2 there). [These errors were later 
corrected in an erratum, which, however, did not 
correct Eq.~(27), nor Fig.~2.]
We shall interpret below the strong sensitivity of $\mu_\ell$ and
$\sigma_\ell$ to $c$, and contrast the vanishing of $\mu_\ell$ and
$\sigma_\ell$ in the formal ``black-hole limit'' $c\to 1/2$, 
to the nonvanishing of the ``shape'' Love numbers $h_\ell$ in the same
limit. In order to approach the ``black-hole limit'' (which is, however,
disconnected from the perfect-fluid star models), we shall particularly
focus on the incompressible models which can reach the maximum compactness
of fluid models, namely $c_{\rm max}=4/9$.

In subsequent works, we shall show how to incorporate the knowledge acquired
here on the various tidal responses of neutron stars into the Effective One
Body (EOB) framework. Indeed, recent 
investigations~\cite{Damour:2009kr,Buonanno:2009qa} have shown
that the EOB formalism is the most accurate theoretical way of 
describing the motion and radiation of inspiralling compact binaries.

This paper is organized as follows: Sec.~\ref{sec:sec2} is an introduction to
the various possible tidal responses of a neutron star. Section~\ref{sec:sec3}
discusses the relevant equations to deal with stationary perturbations of
neutron stars that are then used in Sec.~\ref{sec:sec4} and
Sec.~\ref{sec:sec5} to compute the electric-type ($\mu_\ell$) 
and magnetic-type ($\sigma_\ell$) tidal coefficients. Section~\ref{sec:sec6}
is devoted to the computation of the ``shape'' Love numbers $h_\ell$.
Sections~\ref{sec:sec7},\ref{sec:sec8} and~\ref{sec:sec9}
provide explicit numerical results related to $\mu_\ell$, $\sigma_\ell$ and
$h_\ell$ respectively. The concluding section, Sec.~\ref{sec:conclusions}, 
summarizes our main results.

\section{The various tidal responses of a neutron star}
\label{sec:sec2}

Let us first recall that the motion and radiation of a system of well
separated, strongly self-gravitating (``compact''), bodies can be
theoretically investigated by a ``matching'' approach which consists 
in splitting the problem into two subproblems:

(i) the outer problem where one solves field equations in which the bodies 
are ``skeletonized'' by worldlines endowed with some global characteristics
(such as mass, spin or higher-multipole moments), and

(ii) the inner problem where one obtains the near-worldline behavior of the
outer solution from a study of the influence of the other bodies on the
structure of the fields in an inner world tube around each body.

This matching approach has been used: to obtain the dynamics of binary black
holes at low post-Newtonian
orders~\cite{D'Eath:1975qs,D'Eath:1975vw,Thorne:1984mz},
to prove that the tidal deformation of compact bodies will start to introduce
in the outer problem a dependence on the internal structure of the
constituent bodies (measured by a ``relativistic generalization of the second
Love number'' $k$) only at the fifth post-Newtonian (5PN) 
level~\cite{Damour_LesHouches}, and to derive the dynamics of compact bodies
in alternative theories of gravitation~\cite{Eardley:1975,Will:1981cz,Damour:1992we}.
Finite-size corrections to the leading ``skeletonized'' dynamics can be taken
into account by adding nonminimal worldline couplings to the effective
action~\cite{Damour:1998jk,Goldberger:2004jt}.

Let us start by considering the ``inner problem'' for a neutron star, i.e.,
the influence of the other bodies in the considered gravitationally
interacting system~\footnote{In the following, we shall have in mind a binary
system made either of two neutron stars or of a neutron star and a black
hole.}. As explained, e.g. in Ref.~\cite{Damour_LesHouches}, the matching
method uses a multi-chart approach which combines the information contained in
several expansions. One uses both a global weak-field expansion 
$g_{\mu\nu}(x) = \eta_{\mu\nu} + h_{\mu\nu}^{(1)}(x) +h_{\mu\nu}^{(2)}(x)+\dots$ 
for the outer problem, and several local expansions of the type
\begin{equation}
\label{eq:G_expansion}
G^A_{\alpha\beta}\left(X^\gamma_A\right)=G^{(0)}_{\alpha\beta}\left(X^\gamma_A\right)
+ H_{\alpha\beta}^{(1)}\left(X^\gamma_A\right) + \dots 
\end{equation}
for each inner problem. Here, $G^{(0)}_{\alpha\beta}$ denotes the metric
generated by an isolated neutron star, as seen in a local inner coordinate
system $X^\alpha_A$, which is nonlinearly related to the global
(``barycentric'') coordinate system $x^\mu$ by an expansion of the form
\begin{equation}
x^\mu = z^\mu_A(X^0_A) + e^\mu_a\left(X^0_A\right)X^a_A + \dfrac{1}{2} 
f^\mu_{ab}\left(X^0_A\right)X^a_A X^b_A + \dots
\end{equation}
Here, the suffix $A=1,2,\dots,N$ labels the considered member of the $N$-body
system, while $H^{(1)}_{\alpha\beta}\left(X_A^\gamma\right)$ denotes the
metric perturbation, seen in the local $A$-frame, Bacause of the  combined
influence of the various companions $B\neq A$ of $A$.
In the leading approximation $H^{(1)}_{\alpha\beta}$ is a sum of separate
contributions due to each $B\neq A$: each contribution then contains both the 
far-away field generated by the $B$ worldline, its deformation as it
propagates on the ``background'' metric $G^{(0)}_{\alpha\beta}$ generated by
$A$, and the tidally-induced effect of the deformation of $A$ by the effect of
$B$.

Before tackling the technical problem of computing $H^{(1)}_{\alpha\beta}$,
let us recall the general structure of tidal expansions in 
general relativity~\cite{Thorne:1984mz,Damour:1990pi,Damour:1991yw}.
We will use here the notation and results of the general multi-chart approach
to the general relativistic dynamics of $N$ self-gravitating, deformable
bodies developed by Damour, Soffel 
and Xu (DSX)~\cite{Damour:1990pi,Damour:1991yw,Damour:1992qi,Damour:1993zn}.

Using the DSX notation (with $T\equiv X^0_A/c$),
\begin{align}
G_{00}^A(X)& =-\exp\left(-2W^A/c^2\right),\\
G_{0a}^A(X)& = -\dfrac{4}{c^3} W^A_a, \\
E^A_a(X)   & = \de_a W^A + \dfrac{4}{c^2} \de_T W^A_a \\
B^A_a(X)   & = \epsilon_{abc}\de_b \left(-4 W^A_c\right),
\end{align}
one defines, in the local frame of each body $A$, two sets of
``gravito-electric'' and ``gravito-magnetic'' relativistic tidal moments,
$G_L^A$ and $H_L^A$, respectively as\footnote{As in DSX, $L$ denotes a a
  multi-index $a_1 a_2 \dots a_\ell$  and $\left<a_1 \dots a_\ell\right>$
a symmetric-trace-free (symmetric-trace-free) projection.}
\begin{align}
\label{GL}
G_L^A(T)&\equiv \left[\de_{\left< L-1\right.  }\bar{E}^A_{\left.a_\ell\right>}(T,\X)\right]_{X^a\to 0},\\
\label{HL}
H_L^A(T)&\equiv \left[\de_{\left< L-1 \right .}\bar{B}^A_{\left.a_\ell\right>}(T,\X)\right]_{X^a\to 0},
\end{align}
where $\bar{E}_a^A$ and $\bar{B}_a^A$ denote the externally-generated parts of
the local gravito-electric and gravito-magnetic fields $E^A_a$ and $B^A_a$. 
In the presently considered approximation where $G^{A(0)}_{\alpha\beta}$ is
stationary, and where it is enough to consider the linearized, multipole
expanded, perturbation 
$H^{(1)}_{\alpha\beta}$ in Eq.~\eqref{eq:G_expansion}, the externally
generated parts $\bar{E}^A_a$ and $\bar{B}^A_a$ are well-defined and capture
the terms in $E^A_a$ and $B^A_a$ that asymptotically
grow as $R^{\ell-1}$ as $R\equiv|\X|\to\infty$. The (seemingly contradictory)
formal limit $X^a\to 0$ indicated in Eqs.~\eqref{GL}-\eqref{HL} refers to the
matching performed in the outer problem (where, roughly speaking, the outer
limit $X^a_{\rm outer}\to 0$ can still refer to a worldtube which is large, 
in internal units, compared to the radius of body $A$).

Besides the externally-generated ``tidal moments''\eqref{GL}-\eqref{HL}, one
also defines the internally-generated ``multipole moments'' of body $A$,
$M^A_L(T)$ (mass moments) and $S^A_L(T)$ (spin moments) as the symmetric-trace-free tensors
that parametrize the body-generated terms in the metric coefficients 
$W^A$, $W^A_a$ that asymptotically decrease (in the $A$-body zone) as 
$R^{-(\ell+1)}$ as $R\equiv |\X|\to\infty$. The normalization of these
quantities is defined by Eqs.~(6.9) of~\cite{Damour:1990pi} (and agrees with
the usual one in post-Newtonian theory).

In the stationary case (which is relevant to our 
present ``adiabatic'' approach
to tidal effects), this normalization is such that the 
``internally-generated'' post-Newtonian metric potentials
$W^{+A}(X)$, $W^{+A}_a(X)$ read
\begin{align}
\label{w+}
W^{+A}(X)   &= G\sum_{\ell \geq 0}\dfrac{(-)^{\ell}}{\ell!}\de_L\left(\dfrac{M^A_L}{R}\right),\\
\label{w+a}
W^{+A}_a(X) &=-G\sum_{\ell \geq 1}
\dfrac{\ell}{\ell+1}\dfrac{(-)^\ell}{\ell!}\epsilon_{abc}\de_{bL-1}\left(\dfrac{S^A_{cL-1}}{R}\right)\nonumber\\
&\quad-\dfrac{1}{4}\de_a\left(\Lambda^A-\lambda\right),
\end{align}
where $\Lambda^A-\lambda$ is a gauge transformation (which would drop out if we
had considered the gravito-magnetic field $B^{+A}_a$).

In the first post-Newtonian approximation considered by DSX, the separation
of the local-frame potential $W^A_\alpha (X)$ into an ``externally-generated''
part $\bar{W}^A_\alpha$ and an ``internally-generated'' one $W^{+A}_\alpha$,
is well defined (thanks to the structure of Einstein's equations). In the case
we are considering here of a linearly perturbed, quasi-stationary, fully
relativistic neutron star, the asymptotically growing character (as
$R\to\infty$) of the externally-generated potentials allows one to uniquely
define the tidal moments~\eqref{GL}-\eqref{HL}. On the other hand, the
asymptotic decrease $\propto R^{-(\ell+1)}$ of the internally 
generated  multipolar potentials~\eqref{w+}-\eqref{w+a} introduces an ambiguity
in their definition. For  an attempt to uniquely define the gravito-electric
quadrupole moment $M^A_{ab}$ induced on a black hole by an external tidal
moment $G^A_{ab}$ see~\cite{suen:1986}. Here, instead of relying on such a
conventional (harmonic-coordinates related) definition of the induced multipole
moments, we shall follow the spirit of Sec.~5 of \cite{Damour_LesHouches}
in defining $M^A_L$, $S^A_L$ as parametrizing the (uniquely defined) pieces in
the local-frame metric $G^A_{\alpha\beta}(X)$ which violate the 
``effacing principle'', in that they directly depend on the body $A$ being a
neutron star, rather than a black hole. Reference~\cite{Damour_LesHouches}
explicitly treated the dominant even-parity case, and introduced (see Eq.~(11)
there) a ``dimensionless constant $k$'' ($\equiv a_2$ as defined below) as a
``relativistic generalization of the second Love number''.
This minimal definition (which will be
made fully precise below) is rather natural, and coincides with the definition
adopted in~\cite{Flanagan:2007ix,Hinderer:2007mb}.

With this notation in hands, we can define the two ``tidal-polarizability''
coefficients $\mu_\ell$ and $\sigma_\ell$ introduced 
in Eqs.~(6.19) of~\cite{Damour:1991yw}.
These coefficients relate the (electric or magnetic) tidal
induced~\footnote{Here, we consider a nonrotating star which is spherically
  symmetric (with vanishing multipole moments) when it is isolated, so that
$M^A_L$ and $S^A_L$ represent the multipole moments induced by the influence
of the external tidal fields $G^A_L$ and $H^A_L$.} multipole moments to the 
corresponding external tidal moments, i.e.
\begin{align}
\label{def_mu}
M^A_L &= \mu^A_\ell G^A_L,\\
\label{def_sigma}
S^A_L &= \sigma^A_\ell H^A_L.
\end{align}
The electric-type (or ``even-parity'') tidal coefficient $\mu_\ell$ generalizes
the ($k_{\ell}$-type) Newtonian ``Love number''. For the leading quadrupolar
tide, $\mu_2$, as defined by Eq.~\eqref{def_mu}, agrees with the quantity
denoted $\lambda$ in~\cite{Flanagan:2007ix,Hinderer:2007mb}.
The magnetic-type (or ``odd-parity'') quadrupolar tidal coefficient $\sigma_2$
is proportional to the quantity $\gamma$ which has 
been considered in the investigations of Favata~\cite{Favata:2005da} 
which were, however, limited to the first
post-Newtonian approximation. Here, we shall consider the case of strongly
self-gravitating bodies (neutron stars), and study the dependence of both
$\mu_\ell^A$ and $\sigma_\ell^A$ on the compactness $c_A\equiv (GM/c^2R)_A$ of
the considered neutron star.
Let us also note that, in terms of finite-size corrections to the leading
point-particle effective action $S_{\rm point mass} = -\sum_A \int M_A ds_A$,
the two tidal effects parametrized by $\mu_\ell$ and $\sigma_\ell$ correspond
to nonminimal worldline couplings respectively proportional to
\begin{equation} 
\label{action_tidal}
\mu_\ell^A\int ds_A \left(G^A_L\right)^2,\qquad \text{and}\qquad
\sigma_\ell^A\int ds_A\left(H^A_L\right)^2.
\end{equation}
The leading, quadrupolar corrections \eqref{action_tidal} can be reproduced
(using the link between $G^A_{ab}$ and 
$\u_A^\mu\u_A^\nu\bar{R}^A_{\mu a\nu b}$,
and $H^A_{ab}$ and $\epsilon_b^{\;cd}\u^\mu_A\bar{R}^A_{\mu a c d}$, see
Sec.~3.D of~\cite{Damour:1990pi} 
and~\cite{Thorne:1984mz}) as the following
nonminimal couplings involving the Weyl tensor
\begin{equation}
\label{nonminimal}
\mu_2\int ds\E_{\alpha\beta}\E^{\alpha\beta}\quad\text{and}\quad 
\sigma_2 \int ds \B_{\alpha\beta}\B^{\alpha\beta},
\end{equation}
where \hbox{$u^\mu=dz^\mu/ds$}, and we have introduced the tensors $\E_{\alpha\beta}\equiv u^\mu u^\nu C_{\mu\alpha\nu\beta}$ 
and \hbox{$\B_{\alpha\beta}\equiv u^\mu u^\nu C^*_{\mu\alpha\nu\beta}$}, with
$C^*_{\mu\nu\alpha\beta}\equiv\frac{1}{2}\epsilon_{\mu\nu}^{\;\;\rho\sigma}C_{\rho\sigma\alpha\beta}$
being the dual of the Weyl tensor. In $D=3+1$ dimensions, and in absence of
parity-violating couplings, the two terms~\eqref{nonminimal} are the only
possible isotropic couplings. In higher dimensions, there are three nonminimal
isotropic couplings quadratic in the Weyl tensor as indicated in Eq.(90)
of~\cite{Goldberger:2004jt}. Note that we are using here the freedom of
locally redefining the dynamical variables to eliminate terms proportional to
the (zeroth-order) equations of motion, such as terms involving the Ricci tensor;
see, e.g., the discussion of finite-size effects in tensor-scalar gravity in
Appendix A of Ref.~\cite{Damour:1998jk}.

Let us finally note that there are other ``tidal coefficients'' which might be
interesting to discuss. First, though the linear relations~\eqref{def_mu}-\eqref{def_sigma}
are the most general ones that can exist in the (parity-preserving) case of a
nonspinning neutron star, the tidal properties of a spinning neutron star will
involve other tidal coefficients, proportional to the spin, and associated to
a mixing between electric and magnetic effects. Such electric-magnetic mixing
terms would correspond, say in the leading quadrupolar case, to nonminimal
worldline couplings quadratic in $C_{\mu\nu\alpha\beta}$ and linear in the spin
tensor $S^A_{\mu\nu}$.

There exist also other tidal coefficients which do not have a direct dynamical
meaning, but which generalize the ``first type'' of Love numbers introduced in
the theory of Newtonian tides. Indeed, it is physically meaningful to define, 
for any $\ell$, a ``shape'' Love number measuring the proportionality between
the external tidal influence, and the deformation of the geometry of the
surface of the considered (neutron) star. More precisely, limiting ourselves
to the electric-type tides, one can define a dimensionless number $h_{\ell}$
by writing, as one does in Newtonian theory,
\begin{equation}
g \left(\delta R\right)_\ell = h_\ell U^{\rm disturb}_\ell(R) ,
\ee
or, equivalently,
\be
\label{eq:def_hl}
\left(\dfrac{\delta R}{R}\right)_\ell = h_\ell\dfrac{U_\ell^{\rm disturb}(R)}{g R},
\ee
where $(\delta R/R)_\ell\propto P_\ell(\cos\theta)$ represents the fractional
deformation of the (areal) radius $R$ of the neutron star (measured in a
geometrically invariant way, by relating it to the inner geometry of the
deformed surface), where $U^{\rm disturb}_\ell (R)\propto R^\ell P_\ell(\cos\theta)$
represents the usual, external, Newtonian tidal potential deforming the star,
formally evaluated at the radius of the star (as if one were in flat space),
and where $g\equiv GM/R^2$ represents the usual Newtonian surface gravity of
the neutron star. This $h_\ell$, ``shape'' Love number  has been recently
considered in the theory of the gravitational polarizability of black 
holes~\cite{DL09} and it will be interesting to compare and contrast the
values of the $h_\ell$ for black holes to the values of $h_\ell$ for neutron
stars, especially in the limit where the compactness gets large. See 
Section~\ref{sec:sec6}
below which will give the exact definition of the quantity
$\left(\delta R/R\right)_\ell$.

\section{Stationary perturbations of a neutron star}
\label{sec:sec3}

The unperturbed structure of an isolated (nonrotating) neutron star is
described by a metric of the form
\begin{equation}
\label{eq:bckg}
G^{A(0)}_{\alpha\beta}dX^{\alpha}dX^{\beta}= -e^{\nu(r)}dt^2 + e^{\lambda(r)} dr^2 + r^2d\Omega^2.
\end{equation}
Here, and in the following, for notational simplicity we shall denote the
local (spherical) coordinates of the $A$-body frame simply as
$(t,r,\theta,\varphi)$ 
(with $d\Omega^2\equiv d\theta^2+\sin^2 \theta d \varphi^2$), 
instead of the upper case letters $(T,R,\Theta,\Phi)$ that would more
closely follow the DSX notation recalled above. Introducing as usual the
radial dependent mass parameter $m(r)$ by\footnote{Henceforth, we shall often
  set $G=c=1$}
\be
e^{\lambda(r)}\equiv \left(1-\dfrac{2m(r)}{r}\right)^{-1} ,
\ee
and assuming a perfect-fluid energy-momentum tensor
\begin{equation}
\label{eq:stress_energy}
T_{\mu\nu} = (e+p)u_{\mu} u_{\nu} + p g_{\mu\nu},
\end{equation}
the spherically symmetric metric coefficients $\nu(r)$, $m(r)$ and
the pressure $p(r)$ satisfy the Tolman-Oppenheimer-Volkoff (TOV) equations 
of stellar equilibrium
\begin{align}
\label{tov:1}
\dfrac{d m}{dr}&=4\pi r^2 e ,\\
\label{tov:2}
\dfrac{d p}{dr}&= - (e+p)\dfrac{m+4\pi r^3 p}{r^2 - 2 m r},\\
\label{tov:3}
\dfrac{d \nu}{dr}&= \dfrac{2(m+4\pi r^3 p)}{r^2 - 2 m r} .
\end{align}
These equations are integrated 
from the center outward once that a barotropic EOS 
relating $p$ to $e$ is provided.
We shall consider several types of barotropic EOS, namely two different types
of ``polytropic EOS'' (an $e$-polytrope, with $p=\kappa e^\gamma$, and a
$\mu$-polytrope'', with $p=k\mu^\gamma$ and  $e=\mu + p/(\gamma-1)$, where
$\mu = n m_b$ is the baryonic rest-mass density), and two different
tabulated (``realistic'') EOS (the FPS one~\cite{fps} and the SLy~\cite{sly} one).
In view of the current large uncertainty in the correct description of dense
nuclear matter, we are not claiming that our selection of ``realistic'' EOS
is physically preferred (see, e.g., Ref.~\cite{Bernuzzi:2008fu} and
references therein for a thorough comparison among models from various EOS). 
We have chosen them because they have been used in recent numerical 
relativity simulations of binary neutron star 
coalescence~\cite{Shibata:2005ss,Shibata:2005xz}.
As for the polytropic EOS, they have also been often used in numerical
relativity simulations (especially the $\mu$-polytrope one), and their
dependence on the adiabatic index\footnote{As is well-known, 
the dependence on the ``polytropic constant'' $\kappa$ can be 
absorbed in the definition of suitable 
``polytropic units''.} is a convenient way of varying the
``stiffness'' of the EOS
(the limit $\gamma\to\infty$ representing the stiffest 
possible EOS, namely incompressible matter 
with $e=\text{const}$ and an infinite speed of sound).

Bacause of the  spherical symmetry of the background, the metric perturbation
\begin{equation}
G^{A}_{\alpha\beta}(X) = G^{A(0)}_{\alpha\beta}(X) + H_{\alpha\beta}(X),
\end{equation}
here considered at the linearized level, can be 
analyzed in (tensor) spherical harmonics. 
The metric is expanded in even-parity and 
odd-parity tensor harmonics as 
\begin{equation}
H_{\alpha\beta} = H_{\alpha\beta}^{(\rm e)} + H_{\alpha\beta}^{(o)}.
\end{equation}
In the Regge-Wheeler gauge, and following standard definitions for
the expansion coefficients and the sign conventions 
of~\cite{IP91,Lindblom:1997un}, one has
\begin{align}
\label{metric_even}
H_{\alpha\beta}^{(\rm e)}dX^\alpha dX^\beta & =
 - \big[e^{\nu}H_0^{\ell m}dt^2 + 2H_1^{\ell m} dt dr \nonumber\\
   & + H_2^{\ell m} e^{\lambda} dr^2 
   +r^2 K^{\ell m} d\Omega^2\big]Y_{\lm},
\end{align}
while the nonvanishing components of  
$H_{\alpha\beta}^{(\rm o)}$ are $H_{tA}^{(\rm o)}=h_0 \epsilon_A^{\;\;B}\nabla_B
Y_{\lm}$ and $H_{rA}^{(\rm o)}=h_1 \epsilon_A^{\;\;B}\nabla_B Y_{\lm}$
where $(A,B)=(\theta,\varphi)$ and where $\epsilon_A^{\;\; B}$ is the mixed
form of the volume form on the sphere $S^2_r$.

Our aim is then to solve the {\it coupled system} of the perturbed Einstein's
equations, together with the perturbed hydrodynamical equations 
$\nabla^\alpha \delta T_{\alpha\beta}[e,p]=0$, so as to 
describe a star deformed by an external tidal field. 
We shall only consider {\it stationary} perturbations
(``adiabatic tides'').

\subsection{Even-parity, stationary barotropic perturbations}
\label{sec:even}
Even-parity, stationary perturbations of a barotropic star
simplify in that: (i) the metric perturbations reduce to two
functions $H=H_0=H_2$, and $K$ (with $H_1=0$), (ii) the fluid perturbations
are described by the logarithmic enthalpy function $h$, such that $\delta
h=\delta p/(e+p)$, and (iii) the latter logarithmic enthalpy function 
is simply related (in absence of entropy 
perturbation) to the metric function $H$ by
\begin{equation}
\delta h = -\dfrac{1}{2}H .
\end{equation} 
It was then showed by Lindblom, Mendell and Ipser~\cite{Lindblom:1997un} how
to convert the system of first-order radial differential equations relating
$H'$, $K'$, $H$ and $K$ to a single second-order radial differential equation
for the metric variable $H$ 
(such that $H_{00}=-e^\nu HY_{\lm}$)
of the form
\begin{equation}
\label{eq:H}
H'' + C_1 H' + C_0 H =0.
\end{equation}
[As usual, we shall generally
 drop the multipolar index $\ell$ on the various metric perturbations. The
presence of a factor $Y_{\lm}(\theta,\varphi)$, or $P_\ell (\cos\theta)$, in
(or to be added to) the considered metric perturbation is 
also often left implicit.]
Taking the stationary limit ($\omega\to 0$) of the results given in 
Appendix A of~\cite{Lindblom:1997un} (together with the barotropic relation
$\tilde{\delta} U=0$) one gets
\begin{widetext}
\begin{align}
\label{eq:C1}
C_1 &= \dfrac{2}{r} + \dfrac{1}{2}\left(\nu' - \lambda'\right)=
\dfrac{2}{r}+e^{\lambda}\left[\dfrac{2m}{r^2} + 4\pi r(p-e)\right],\\
C_0 &= e^{\lambda}\left[-\dfrac{\ell(\ell+1)}{r^2} + 4\pi(e+p)\dfrac{de}{dp} +
  4\pi (e+p)\right] + \nu'' + \left(\nu'\right)^2
+\dfrac{1}{2r}\left(2-r\nu'\right)\left(3\nu'+\lambda'\right)\nonumber\\
&=e^{\lambda}\left[-\dfrac{\ell(\ell+1)}{r^2} + 4\pi
  (e+p)\dfrac{de}{dp}+4\pi\left(5 e + 9p\right)\right]-(\nu')^2,
\label{eq:C0}
\end{align}
\end{widetext}
where we have used the background (TOV) equations to rewrite $C_1$ and
$C_0$. As a check, we have also derived from scratch Eq.~\eqref{eq:H}
by starting from the ``gauge-invariant'' formalism of
Ref.~\cite{Gundlach:1999bt}.
Equation~\eqref{eq:H} generalizes to an arbitrary value of 
the multipolar order $\ell$ Eq.~(15) of 
Ref.~\cite{Hinderer:2007mb}, which concerned the leading
quadrupolar even-parity tide.

For completeness, let us note that the other metric variable, $K$, can be
expressed as a linear combination of $H$ and $H'$, namely
\begin{equation}
K = \alpha_1 H' + \alpha_2 H ,
\end{equation}
where the explicit expressions of the coefficients $\alpha_1$ and $\alpha_2$
can also be deduced by taking the stationary limit of the results given in
Appendix A of~\cite{Lindblom:1997un}.

\subsection{Odd-parity, stationary perturbations}
\label{sec:odd}

It was shown by Thorne and Campolattaro~\cite{TC67} 
that odd-parity perturbations of
a nonrotating perfect-fluid star consists only of metric fluctuations, and do
not affect the star's energy density and pressure. One might naively think
that this means that an odd-parity tidal field will induce no
(gauge-invariant) spin multipole moment in a (nonrotating) star. This
conclusion is, however, incorrect because the ``gravitational potential well''
generated by the stress-energy tensor of the star does affect the ``radial
propagation'' of the external odd-parity tidal fields and necessarily adds an
asymptotically decreasing ``induced'' tidal response to the ``incoming'' tidal
field. To describe this phenomenon, it is convenient to describe the
odd-parity perturbation by means of the (static limit of the) 
``master equation'' derived by  Cunningham, Price and 
Moncrief~\cite{CPMI} (see also Ref.~\cite{Andrade:1999mj}). 
In the stationary limit, and in terms of the ordinary radial variable $r$ 
(rather than the ``tortoise'' coordinate $r_*$) this equation reads
\begin{widetext}
\begin{equation}
\label{eq:odd_master}
\psi'' +
\dfrac{e^{\lambda}}{r^2}\left[2m+4\pi r^3(p-e)\right]\psi' 
- e^{\lambda}\left[\dfrac{\ell(\ell+1)}{r^2}-\dfrac{6m}{r^3}+4\pi(e-p)\right]\psi=0.
\end{equation}
\end{widetext}
In terms of the variables $(h_0,h_1)$ entering the odd-parity perturbations,
the odd-parity master function $\psi$ can be taken to be either
$e^{(\nu-\lambda)/2}h_1/r$, or the combination $r\de_t h_1-r^3\de_r(h_0/r^2)$
[see e.g.~\cite{Nagar:2005ea} for more details]. As $h_1$ vanishes in the 
stationary limit, we can define $\psi$ as being
\begin{equation}
\label{def:psi}
\psi = r^3\de_r\left(\dfrac{h_0}{r^2}\right)=rh_0'-2h_0.
\end{equation}

\section{Computation of the electric-type 
tidal coefficient $\boldsymbol{\mu_\ell}$}
\label{sec:sec4}

The electric-type tidal response coefficient $\mu_\ell$, defined by
Eq.~\eqref{def_mu} above, can be obtained by going through three
steps: (i) numerically solving the even-parity master equation~\eqref{eq:H}
within the neutron star; (ii) analytically solving the same 
master equation~\eqref{eq:H} in the exterior of the star; and (iii) matching
the interior and exterior solutions across the star surface, taking into
account the definition~\eqref{def_mu} to normalize the ratio between the
``growing'' and ``decreasing'' parts of $H(r)$, namely $H^{\rm growing}\sim
r^\ell$ versus $H^{\rm decreasing}\sim \mu_\ell r^{-(\ell+1)}$.

\subsection{The internal problem}
\label{sec:internal}

The internal value of the metric function $H(r)$ is obtained by numerically
integrating Eq.~\eqref{eq:H}, together with the TOV
equations~\eqref{tov:1}-\eqref{tov:3},
from the center  (or, rather some very small cut-off radius $r_{0 }=10^{-6}$)
outwards, starting with some central values of $m,p,\nu,H$ and $H'$. 
For $H$, one takes as starting values at the cut-off radius 
$H(r_{0})=r^\ell_0$ and $H'(r_0)=\ell r_0^{\ell -1}$.
The latter boundary conditions follow from the analysis of Eq.~\eqref{eq:H}
around the regular-singular point $r=0$, which shows that 
$H(r)\simeq \bar{h} r^\ell$ (where $\bar{h}$ is an 
arbitrary constant) is the most general regular solution around $r=0$.
As Eq.~\eqref{eq:H} is homogeneous in $H$, 
the scaling constant $\bar{h}$ is irrelevant and will 
drop out when we shall match the logarithmic derivative
\begin{equation}
\label{eq:yint}
y^{\rm int}(r)\equiv \dfrac{r H'}{H},
\end{equation}
across the star surface. This is why it is enough to use $\bar{h}=1$ 
as initial boundary conditions for $H$.

The main output of this internal integration procedure is to compute (for each
value of $\ell$) the value of the internal logarithmic
derivative~\eqref{eq:yint} at the star's surface, say $r=R$
\begin{equation}
y_\ell \equiv y_{\ell}^{\rm int}(R) .
\end{equation}

\subsection{The external problem}
\label{eq:external}

As noticed long ago by Regge and Wheeler~\cite{Regge:1957td} 
and Zerilli~\cite{Zerilli:1971wd}, the exterior form of the stationary, 
even-parity master equation~\eqref{eq:H} ($e=p=0$, $m(r)=M$) can be 
recast as an associated Legendre equation (with $\ell=\ell$ and $m=2$). 
More precisely, in terms of the independent variable $x\equiv r/M-1$, 
the exterior form of~\eqref{eq:H} reads 
\begin{equation}
(x^2 -1) H'' + 2x H' - \left(\ell(\ell+1) + \dfrac{4}{x^2
    -1}\right)H=0 ,
\end{equation}
where the prime stands now for $d/dx$. Its general solution can be written as
\begin{equation}
\label{H_external}
H = a_P \hat{P}_{\ell 2}(x) + a_Q \hat{Q}_{\ell 2}(x),
\end{equation}
where the hat indicates that the associated Legendre functions of 
first, $P_{\ell 2}$, and second\footnote{Note that, contrary to the usual
  mathematical definition of $Q_{\lm}(x)$, which is tuned to the real interval
$-1<x<+1$, we need to work with $Q_{\lm}(x)$ in the interval $x>1$, This means
replacing $\log\left(\dfrac{1+x}{1-x}\right)$ with 
$\log\left(\dfrac{x+1}{x-1}\right)$.}, $Q_{\ell 2}$, kind  have been 
normalized so that $\hat{Q}_{\ell 2}\simeq 1/x^{\ell+1}\simeq (M/r)^{\ell+1}$
and $\hat{P}_{\ell2}\simeq x^\ell \simeq (r/M)^{\ell}$ when $x\to\infty$ or 
$r\to\infty$; $a_Q$ and $a_P$ are integration constants to be determined 
by matching to the internal solution. Defining 
$a_{\ell}\equiv a_Q/a_P$, the exterior logarithmic derivative 
$y^{\rm  ext}\equiv rH'/H$ reads 
\begin{equation}
\label{eq:external_y}
y_{\ell}^{\rm ext}(x) = (1+x)\dfrac{\hat{P}_{\ell2}'(x) + a_{\ell}\hat{Q}_{\ell2}'(x)}
{\hat{P}_{\ell2}(x) + a_{\ell}\hat{Q}_{\ell2}(x) }.
\end{equation}

\subsection{Matching at the star's surface, and computation of the 
``electric'' tidal Love number}

As Eq.~\eqref{eq:H} is second-order in the radial derivative of $H$, one
expects that $H$ and $H'$ will be continuous at the star's surface. Actually,
the issue of regularity at the star surface is somewhat subtle because some of
the thermodynamic variables (such as pressure) do not admit regular Taylor
expansions in $r-R$ as $r\to R$. For instance, while the logarithmic enthalpy 
$h(p)=\int_0^p dp/(e+p)$ vanishes smoothly ($h(r)\propto r-R$)
across the surface, one finds that (for any polytrope) 
$p(r)\propto (r-R)^{\gamma/(\gamma-1)}$ and that the term involving the inverse
of the squared sound velocity $c_s^2 = dp/de$ in Eq.~\eqref{eq:H} is singular
(when $\gamma>2$), namely
\begin{equation}
(e + p)\dfrac{de}{dp}\propto (r-R)^{\frac{2-\gamma}{\gamma-1}}.
\end{equation}
Despite this mildly singular behavior
of  the coefficient $C_0$ of~\eqref{eq:H} and despite the fact that the exact
location of the tidally-deformed star surface is slightly displaced from the
``background'' value $r=R$, one checks that it is correct 
(when $\gamma<\infty$) to impose the continuity of 
$H$ and $H'$ at $r=R$. 
[Note that we consider here the  case of a finite adiabatic
index $\gamma$. The incompressible limit $\gamma\to\infty$ 
leads to a master equation which is singular at the surface, 
and which must be considered with care. See below our discussion 
of the incompressible limit.]
This continuity then imposes the continuity of the logarithmic 
derivative $r H'/H$. This leads to the condition 
$y^{\rm ext}(R)=y^{\rm int}(R)=y_\ell$, which determines the 
value of the ratio $a_\ell=a_Q/a_P$ in terms of the
compactness $c\equiv M/R$ of the star
\begin{equation}
\label{eq:a_ell}
a_{\ell} = -\left.\dfrac{\hat{P}_{\ell 2}'(x) - c y_{\ell}
\hat{P}_{\ell2}(x)}{\hat{Q}_{\ell 2}'(x) - c y_{\ell}
\hat{Q}_{\ell2}(x)  }\right\vert_{x=1/c-1}.
\end{equation}
On the other hand, the ratio $a_\ell\equiv a_Q/a_P$ can be related to the
tidal coefficient $\mu_\ell$ by comparing (modulo an overall factor $-2$),
\begin{align}
-&\left(\delta H_{00}e^{-\nu}\right)^{\rm growing} = H^{\rm
    growing}(r)\nonumber\\
&\hspace{1.2cm}=a_P
\hat{P}_{\ell 2}(x) Y_{\lm}\simeq a_P\left(\dfrac{r}{M}\right)^{\ell}Y_{\lm},\\
-&\left(\delta H_{00}e^{-\nu}\right)^{\rm decreasing} = H^{\rm
  decreasing}(r)\nonumber\\
&\hspace{1.2cm} = a_Q \hat{Q}_{\ell2}(x)Y_{\lm}\simeq a_Q
\left(\dfrac{r}{M}\right)^{-(\ell+1)}Y_{\lm},
\end{align}
respectively to
\begin{align}
\bar{W}&=\dfrac{1}{\ell!} \hat{X}^LG^A_L=\dfrac{1}{\ell!} r^{\ell} \hat{n}^L
G^A_L ,\\
W^+&=G\dfrac{(-)^\ell}{\ell!}\de_L\left(\dfrac{M^A_L}{r}\right),
\end{align}
(see e.g., Eq.~(4.15a) of Ref.~\cite{Damour:1991yw}) where $n^a\equiv X^a/r$
is a radial unit vector.
Using the fact that
\be
\de_L r^{-1} = (-)^\ell (2\ell-1)!!\,\hat{n}^L \,r^{-(\ell+1)},
\ee
and $M_L=\mu_\ell G_L$, and remembering that 
$G_L\hat{n}^L\propto Y_{\lm}(\theta,\varphi)$, we see that
\begin{equation}
(2\ell-1)!!G\mu_\ell = \dfrac{a_Q}{a_P}\left(\dfrac{GM}{c_0^2}\right)^{2\ell+1}
=a_\ell \left(\dfrac{GM}{c_0^2}\right)^{2\ell+1}.
\end{equation}
Note that $G\mu_\ell$ has the dimensions of $[\text{length}]^{2\ell+1}$. There
are then two natural ways of expressing $G\mu_\ell$ in terms of a
dimensionless quantity. Either by scaling it by the $(2\ell+1)$-th
power of $GM/c_0^2$, which leads to
\begin{equation}
(2\ell-1)!! \dfrac{G\mu_\ell}{(GM/c_0^2)^{2\ell+1}}=a_\ell ,
\end{equation}
or by scaling it by the $(2\ell+1)$-th power of the star radius $R$, 
which gives
\be
\label{eq:def_kl}
(2\ell-1)!!\dfrac{G\mu_\ell}{R^{2\ell+1}}\equiv 2k_\ell=a_\ell c^{2\ell+1}.
\ee
Alternatively, we can write
\begin{equation}
G\mu_\ell =
\dfrac{a_\ell}{(2\ell-1)!!}\left(\dfrac{GM}{c_0^2}\right)^{2\ell+1}
=\dfrac{2k_\ell}{(2\ell-1)!!}R^{2\ell+1} .
\end{equation}
The scaling of $G\mu_\ell$ by means of $R^{2\ell+1}$ is the traditional
``Newtonian'' way of proceeding, and leads to the introduction of the
dimensionless ``second tidal Love number'' $k_\ell$ (conventionally
normalized as in Eq.~\eqref{eq:def_kl} above).

One can finally write $k_\ell$ as
\begin{equation}
\label{eq:kl}
k_\ell = \dfrac{1}{2}c^{2\ell+1}a_\ell = 
-\frac{1}{2}\,c^{2\ell +1}\,\left.\dfrac{\hat{P}_{\ell 2}'(x)-c y_{\ell} \hat{P}_{\ell 2}(x)}
{\hat{Q}_{\ell 2}'(x)- c y_{\ell} \hat{Q}_{\ell 2}(x)}\right\vert_{x=1/c-1}.
\end{equation}
The dimensionless Love number $k_\ell$ has the advantage of having a weaker
sensitivity on the compactness $c\equiv GM/(c_0^2R)$ (especially as the
compactness formally tends to zero, i.e. in the Newtonian limit). Note,
however, that the dimensionless quantity which will most directly enter 
the gravitational-wave phase of inspiralling binary neutron stars (NS) is
$G\mu_\ell/(GM/c_0^2)^{2\ell +1}\sim a_\ell \sim c^{-(2\ell +1)} k_\ell$.

The evaluation of the result~\eqref{eq:kl} for $k_\ell$ yields the following
explicit expressions for $2\leq \ell\leq 4$ (with, for simplicity, $y\equiv y_\ell$):
\begin{widetext}
\begin{align}
\label{eq:k_ell2}
k_2 &= \frac{8}{5} (1-2 c)^2 c^5 [2 c (y-1)-y+2]\nonumber\\
&\times \bigg\{2 c \left(4 (y+1) c^4+(6 y-4) c^3+(26-22 y) c^2+3 (5 y-8) 
c-3 y+6\right)\nonumber\\
&-3 (1-2 c)^2 (2 c (y-1)-y+2)\log \left(\frac{1}{1-2 c}\right)\bigg\}^{-1},\\
\label{eq:k_ell3}
k_3 &= \frac{8}{7} (1-2 c)^2 c^7 \left[2 (y-1) c^2-3 (y-2)c+y-3\right]\nonumber\\
&\times \bigg\{2 c \left[4 (y+1) c^5+2 (9 y-2) c^4-20 (7 y-9) c^3+5 (37 y-72)c^2-45 (2 y-5) c+15 (y-3)\right]\nonumber\\
& -15 (1-2 c)^2 \left(2 (y-1) c^2-3 (y-2) c+y-3\right) 
\log\left(\frac{1}{1-2 c}\right)\bigg\}^{-1},\\
\label{eq:k_ell4}
k_4 &= \frac{32}{147} (1-2 c)^2 c^9 
\left[12 (y-1) c^3-34 (y-2) c^2+28 (y-3)c-7 (y-4)\right]\nonumber\\
&\times \bigg\{2 c \left[8 (y+1) c^6+(68 y-8) c^5+(1284-996 y) c^4+40 (55 y-116)
   c^3+(5360-1910 y) c^2+105 (7 y-24) c-105 (y-4)\right]\nonumber\\
& -15 (1-2 c)^2 \left[12 (y-1) c^3-34 (y-2) c^2+28 (y-3) c-7 (y-4)\right] 
\log \left(\frac{1}{1-2 c}\right)\bigg\}^{-1}.
\end{align}
\end{widetext}
Equation~\eqref{eq:k_ell2} above agrees with the corrected version
of Eq.~(23) of~\cite{Hinderer:2007mb}.
Note that, independently of the values of $y$ (as long as it does not
introduce a pole singularity, which will be the case), the 
results~\eqref{eq:k_ell2}-\eqref{eq:k_ell4} (and, more generally, the
result~\eqref{eq:kl}) contain an overall factor $(1-2c)^2$ which formally
tends (quadratically) to zero when the compactness $c=GM/R$ ``tends''
toward the compactness of a black hole, namely $c^{\rm BH}=GM/(2 GM)=1/2$.
[The singular logarithm $\log[1/(1-2c)]$ in the denominator is also 
easily checked to be always multiplied by $(1-2c)^2$ and thereby not to affect
the $\propto (1-2c)^2$ formal vanishing of $k_\ell$ as $c\to 1/2$.]
This property can be easily understood as a consequence of the 
``no-hair'' properties of black holes.
Indeed, among the two solutions of the exterior tidal
perturbation equation~\eqref{eq:H}, the no-hair property means that the
solution which is ``rooted'' within the horizon, i.e., the ``asymptotically
decreasing'' solution $Q_{\ell 2}(x)$ is {\it singular} at the horizon,
i.e. when $x=R/M-1\to 1$. More precisely, this singular behavior is
\begin{align}
Q_{\ell 2}(x)&\sim (x^2-1)^{2/2}\dfrac{d^2 Q_\ell (x)}{dx^2}\nonumber\\
&\sim (x^2-1)\dfrac{d^2}{dx^2}
\left[\log\left(\dfrac{x+1}{x-1}\right)P_\ell(x)\right]\nonumber\\
&\sim (x^2-1)(x-1)^{-2}\sim (x-1)^{-1} ,
\end{align}
so that the most singular term in the denominator of $a_\ell$
or $k_\ell$ is $\hat{Q}'_{\ell 2}(x)\sim (x-1)^{-2}\sim (R-2M)^{-2}\sim (1-2c)^{-2}$
which is at the origin of the presence of a factor $(1-2c)^2$ in $a_\ell$
and $k_\ell$. One might naively think that this behavior proves that the
``correct'' value of the $k_\ell$ tidal Love numbers of a black hole is simply
$k_\ell^{\rm BH}=0$. However, we do not think that this conclusion is warranted.
Indeed, as we explained above, the definition used here 
(and in~\cite{Damour_LesHouches,Flanagan:2007ix,Hinderer:2007mb}) 
of the Love numbers of a (neutron) star consists in selecting, 
within the gravitational field of a
tidally distorted star, the terms which violate the ``effacing principle''
(in the sense of Ref.~\cite{Damour_LesHouches}), i.e. the 
internal-structure-dependent terms which differentiate the tidal response of a 
(compact) star, from that of a black hole. From this point of view, the
vanishing of $k_\ell$ as $c\to c^{\rm BH}$ is mainly a consistency check on
this formal definition. The question of computing the ``correct'' value of
$k_\ell$ for a black hole is a technically much harder issue which involves
investigating in detail the many divergent diagrams that enter the computation
of interacting point masses at the 5-loop (or 5PN) level.

Indeed, the issue at stake is the following. When describing the motion of two
black holes (as seen in the ``outer problem'') by a skeletonized action of the
form $S=S_{\rm pointmass} + S_{\rm nonminimal}$, the presence of nonminimal
worldline couplings $S_{\rm nonminimal}$ of the type~\eqref{action_tidal} and~\eqref{nonminimal}
can only be detected if one treats (when using perturbative expansions in
powers of $G$) the general relativistic nonlinear self-interactions 
entailed by $S_{\rm pointmass}=-\sum_A\int M_A ds_A$
at the order of approximation corresponding to $S_{\rm nonminimal}$.
For a black hole (of ``radius'' $R_A= 2GM_A/c_0^2$), the leading nonminimal
coupling parameter scales as $\mu_2^A\sim k_2^A R^5_A/G\sim k_2^A G^4
M_A^5$, so that (using ${\cal E}^A_{\alpha\beta}\sim
R^A_{\alpha\beta\gamma\delta}\propto GM_B$) the leading nonminimal interaction
$\mu_2^A\int ds_A\E^A_{\alpha\beta}\E_A^{\alpha\beta}$ is proportional to
$k^A_2 G^6 M_A^5 M_B^2$. The presence of an overall factor $G^6$ (which is the
same factor $G^6$ that appeared in Eq.~(19) in Sec.~5  of~\cite{Damour_LesHouches}) signals
that such an effect is $G^5$ smaller than the leading (Newtonian) interaction
($\propto G M_A M_B$) between two point masses, so that it corresponds to the
5PN level. In the diagrammatic language of (post-Minkowskian or post-Newtonian)
perturbation theory (as used, e.g., in~\cite{Damour:1995kt}), this corresponds
to the 5-loop level.
Let us recall that the computation of the interaction of two black holes at
the 3-loop level was a technically complex enterprise that necessitated the
careful consideration of many divergent diagrams, and the use of the efficient
method of dimensional regularization~\cite{Damour:2001bu,Blanchet:2003gy}.
At the 3-loop level the result of the computation was (essentially) 
{\it  finite}, though the use of harmonic coordinates in one of the
computations~\cite{Blanchet:2003gy} introduced some gauge-dependent
infinities. As argued long ago~\cite{Damour_LesHouches}, and confirmed by an
effective action approach~\cite{Goldberger:2004jt}, one expects to see real,
gauge-independent infinities arising at 5-loop (5PN), i.e. at the level where
the effacing principle breaks down, and where, as explained above, 
a parameter ($\sim k_2$) linked to the internal structure of the considered 
compact body starts to enter the
dynamics. Until a careful analysis of the 5PN nonlinear self-interactions is
performed, one cannot conclude from the above result ($k_2^{\rm NS}\to 0$ as
$c \to c^{\rm BH}$) that the effective action describing the dynamics of
interacting black holes is described by the pure point-mass action 
$-\sum_A\int m_A ds_A$ without the need of additional nonminimal couplings of
the type of Eq.~\eqref{nonminimal}.

We have phrased here the problem within standard (post-Minkowskian or
post-Newtonian) perturbation theory, because this is the clearest framework
within which the issue of higher order nonlinear gravitational 
interactions of point masses is technically well defined 
(when using, say, dimensional regularization to
define the perturbative interaction of point masses in 
general relativity~\cite{Damour:2001bu,Blanchet:2003gy}).
Note that, in the extreme mass ratio limit ($M_A\ll M_B$), where one might
use black hole perturbation theory, the interaction associated to the leading
nonminimal coupling parameter $\mu_2^A$ of $M_A$ is proportional to $M_A^5$
(see above). This is well beyond the currently studied ``gravitational
self-force'' effects, which are proportional to $M_A^2$, and correspond to 
a ``1-loop'' effect within a black hole background.
\section{Computation of the magnetic-type tidal 
coefficient $\boldsymbol{\sigma_\ell}$}
\label{sec:sec5}

The magnetic-type tidal response coefficient $\sigma_\ell$, defined 
by Eq.~\eqref{def_sigma} above, can be obtained by following three steps, 
which are similar to those followed for the electric-type coefficient $\mu_\ell$.

\subsection{The internal problem}

The internal value of the odd-parity master function $\psi$ is obtained 
by numerically integrating Eq.~\eqref{eq:odd_master}, together with the 
TOV equations. The boundary conditions are now obtained from the behavior
$\psi\propto r^{\ell+1}$ of the general regular solution at the origin. 
Again, the main output of the internal integration procedure is to compute
(for each value of $\ell$) the value of the internal logarithmic derivative 
of $\psi$, at the star surface, say
\begin{equation}
\label{eq:y_odd}
y_\ell^{\rm odd}\equiv y_{\ell}^{\rm int}(R)\equiv 
\left[\dfrac{r\psi_{\rm int}'}{\psi_{\rm int}}\right]_{r=R}.
\end{equation}

\subsection{The external problem}

As noticed long ago by Regge and Wheeler~\cite{Regge:1957td},
the stationary odd-parity perturbations can be analytically solved in the
exterior region. Similar to the even-parity case there exist two types 
of exterior solutions: a ``growing'' type solution, say $\psi_P(\r)$,
with  $\r\equiv r/M$, and a ``decreasing'' type one, say $\psi_Q(\r)$.
We normalize them so that $\psi_P(\r)\simeq \r^{\ell+1}$, and 
 $\psi_Q(\r)\simeq \r^{-\ell}$ as $\r\to\infty$. The general analytical
forms of $\psi_P$ and $\psi_Q$, for any $\ell$, can be obtained 
from Ref.~\cite{Regge:1957td}. In the case of the leading quadrupolar 
odd-parity perturbation, $\ell=2$, the ``growing'' analytical
exterior solution of \eqref{eq:odd_master} is the very simple polynomial
\begin{equation}
\psi_P^{\ell=2}(\r)=\r^3,
\end{equation}
 while the ``decreasing'' one can be expressed in
terms of an hypergeometric function $F(a,b;c;z)$ as
\be
\psi_Q^{\ell=2}(\r)=-\dfrac{1}{4}\r^3\de_\r
\left[\r^{-4}F\left(1,4;6;\dfrac{2}{\r}\right)\right].
\ee
The normalization of $\psi_Q(\r)$ is such that $\psi_Q(\r)\simeq \r^{-2}$ 
as $r\to\infty$. Note also that, for the special values $a=1$, $b=4$, $c=6$,
the hypergeometric function is actually expressible in terms of elementary
functions. The result has the form
\begin{equation}
\label{exp_psiQ}
\psi_Q^{\ell=2}(\r) = A_3 \r^3 \log\left(\dfrac{\r-2}{\r}\right) + A_2 \r^2 + A_1\r +A_0 +
A_{-1} \r^{-1} .
\end{equation}
Going back to an arbitrary $\ell$, the general exterior 
solution of Eq.~\eqref{eq:odd_master} can be written,
analogously to the even-parity Eq.~\eqref{H_external}, as
\be
\psi^{\rm ext} = b_P \psi_P(\r) + b_Q\psi_Q(\r).
\ee
This result allows one to compute the logarithmic derivative 
$y^{\rm odd} = r\psi'/\psi$ of $\psi$ in the exterior domain,
namely
\be
y^{\rm ext}_{\rm odd}(\r)=\hat{r}\dfrac{\psi'_P(\r)+b_\l\psi'_Q(\r)}
{\psi_P(\r)+b_\l\psi_Q(\r)},
\ee
where $b_\l\equiv b_Q/b_P$.

\subsection{Matching at the star surface, and computation of 
the ``magnetic'' tidal  Love number}

We again impose the continuity of $\psi$, $\psi'$, and therefore 
$y^{\rm odd}=r\psi'/\psi$, at the star's surface. Similarly to
the even parity case, this determines the value of the ratio 
$b_\l=b_Q/b_P$ in terms of the compactness of the star:
\be
\label{eq:b}
b_\l = -\left.\dfrac{\psi'_P(\r) - c y_{\rm odd} \psi_P(\r)}
                 {\psi'_Q(\r) - c y_{\rm odd} \psi_Q(\r)}\right|_{\r=1/c}.
\ee
Again, we see at work the effect of the ``no-hair'' property in that the term
$\psi'_Q(\r)$ in the denominator of~\eqref{eq:b} will become (from~\eqref{exp_psiQ}) 
singular as $(\r-2)^{-1}$ when $\r\to 2$. This implies that $b_\l$ will 
vanish proportionally to the first power of $1-2c$ in the formal limit 
where the star's compactness $c\to c^{\rm BH}=1/2$.

The dimensionless quantity $b_\l$, Eq.~\eqref{eq:b}, is the odd-parity analog of
the even-parity quantity $a_\ell$, Eq.~\eqref{eq:a_ell}.
In the even-parity case, $a_\ell$ was, essentially, the tidal response
coefficient $G\mu_\ell$ scaled by $(GM/c_0^2)^{2\ell+1}$. In the present
odd-parity case, the tidal response coefficient $G\sigma_\ell$ has again the 
dimension $[\text{length}]^{2\ell+1}$, and $b_\ell$ (for a general $\ell$) is 
essentially $b_\ell \sim G\sigma_\ell(GM/c_0^2)^{-(2\ell+1)}$. Before working
out the exact numerical coefficient in this proportionality, we can note that
the odd-parity analog of $k_\ell$ (i.e. essentially
$k_\ell\sim G\mu_\ell/R^{2\ell+1}\sim c^{2\ell+1}a_\ell$) will be obtained by
scaling $G\sigma_\ell$ by the \hbox{$(2\ell+1)$-th} power of the star radius
$R$, and will therefore involve the new dimensionless combination
\be
\label{eq:cb}
j_\l\equiv c^{2\ell+1} b_\ell = - c^{2\ell+1}\left.\dfrac{\psi_P'(\r)-cy_{\rm odd}\psi_P(\r)}
                 {\psi'_Q(\r)-cy_{\rm odd}\psi_Q(\r)}\right|_{\r=1/c} .
\ee
One expects that the new odd-parity dimensionless 
combination $j_\l$~\eqref{eq:cb} will, like $k_\ell$, 
depend less strongly on the value of the compactness $c$ 
than $b_\ell$ itself.

Let us now derive the precise link between the odd-parity tidal response
coefficient $G\sigma_\ell$, defined by Eq.~\eqref{def_sigma}, and the
dimensionless quantities $b_\ell$, Eq.~\eqref{eq:b}, or $b_\ell c^{2\ell+1}$,
Eq.~\eqref{eq:cb}.                         
To relate them, we start by noting that the Regge-Wheeler metric function
$h_0$ entering the odd-parity master quantity $\psi$, Eq.~\eqref{def:psi},
parametrizes the $\text{time}\times\text{angle}$ off-diagonal component 
of the metric perturbation 
\be
\label{H0A}
H_{0A}\propto h_0(r)\epsilon_A^{\;\;B}\nabla_BY_{\lm}(\theta,\varphi) ,
\ee
where $A,B=2,3=\theta,\varphi$ are indices on the background coordinate sphere
$S^2_r$ of radius $r$. The metric on $S^2_r$ is $\gamma_{AB}dx^A
dx^B=r^2d\Omega^2$, while $\epsilon_A^{\;\;B}\equiv \gamma^{BC}\epsilon_{AC}$
denotes the mixed form of the volume form
$\frac{1}{2}\epsilon_{AB}dx^A\wedge dx^B=r^2\sin\theta d\theta\wedge d\varphi$ on $S^2_r$.
Let us now consider the gravito-magnetic field $B_a$, as defined by
DSX. Modulo an irrelevant numerical factor, it is the $3$-dimensional curl of
the time-space off-diagonal metric 
component: $B_a\propto \epsilon_{abc}\de_b H_{0c}$.
Let us focus on the ``radial component'' of the gravito-magnetic field $B_a$,
i.e. the pseudo-scalar
\be
{\bf n}\cdot{\bf B} = n^a B_a\propto n^a\epsilon_{abc}\de_bH_{0c}\propto
\epsilon^{AB}\nabla_A H_{0B}
\ee
Using then Eq.~\eqref{H0A}, one finds that
\be
{\bf n}\cdot{\bf B}\propto - h_0(r)\epsilon^{AB}\epsilon_A^{\;\;C}\nabla_B\nabla_C
Y_\lm= \ell(\ell+1)\dfrac{h_0(r)}{r^2}Y_{\lm}, 
\ee
where one used $\epsilon^{AB}\epsilon_A^{\;\;C}= \gamma ^{BC}$ 
 and the fact that
$\gamma^{AB}\nabla_A\nabla_BY_{\lm}=-\ell(\ell+1)r^{-2}Y_{\lm}$,
where the factor $r^{-2}$ comes from the fact that $\gamma_{AB}$ is the metric
on a sphere of radius $r$, rather than a unit sphere. [The various
  proportionality signs refer to irrelevant, coordinate-independent, numerical
factors.] Finally, we have the link
\be
\label{connect_psi_B}
\psi =r^3\de_r\left(\dfrac{h_0}{r^2}\right)\propto r^3 \de_r 
({\bf n}\cdot {\bf B}) .
\ee
Focusing on the two crucial (growing or decreasing) asymptotic terms in the
odd-parity metric, we can now compare the definition of $b_\ell$, namely
\be
\label{psi_prop}
\psi\propto \left\{ r^{\ell+1} 
+ b_\ell \left(\dfrac{GM}{c_0^2}\right)^{2\ell+1}r^{-\ell}\right\}Y_{\lm}(\theta,\varphi),
\ee
to the stationary limit of the general gravito-magnetic fields in a local
$A$-frame (see Eqs.~(2.19) and (4.16) of Ref.~\cite{Damour:1991yw}),
\begin{align}
B_a &= \bar{B}_a + B^+_a \nonumber\\
    &= \sum_\ell \dfrac{1}{\ell!} X^L H_{aL} + \sum_\ell 4G
\dfrac{(-)^\ell}{\ell !} \dfrac{\ell}{\ell+1}\de_{aL}\left(\dfrac{S_L}{r}\right).
\end{align}
Inserting now $S_L=\sigma_\ell H_L$, contracting $B_a$ with $n^a$, 
and recalling that one has $n^a\de_a = \de_r$ and 
\hbox{$\de_L\left( r^{-1}\right)=(-)^\ell(2\ell-1)!! r^{-(\ell+1)}$},
one finds
\begin{align}
{\bf n}\cdot{\bf B} & =\sum_\ell \dfrac{1}{(\ell-1)!}r^{\ell-1}H_L n^L\nonumber\\
&+\sum_\ell 4G\sigma_\ell \dfrac{(-)^\ell}{\ell!}\dfrac{\ell}{\ell+1}\de_r\de_L\left(\dfrac{H_L}{r}\right)\nonumber\\
&= \sum_\ell \dfrac{1}{(\ell-1)!}\left\{r^{\ell-1} - 4G\sigma_\ell (2\ell-1)!!
\dfrac{1}{r^{\ell+2}}\right\}n_LH_L,
\end{align}
so that
\begin{align}
r^3\de_r({\bf n}\cdot{\bf B})&=\sum_\ell\dfrac{1}{(\ell-2)!}\nonumber\\
&\times \left\{r^{\ell+1}
+4G\sigma_\ell(2\ell-1)!!\dfrac{\ell+2}{\ell-1}\dfrac{1}{r^\ell}\right\}n_L H_L.
\end{align}
The comparison with Eq.~\eqref{psi_prop}
finally yields
\begin{align}
\dfrac{4(\ell+2)}{\ell-1}\,(2\ell-1)!! G\sigma_\ell =
b_\ell\left(\dfrac{GM}{c_0^2}\right)^{2\ell+1}
 =b_\ell c^{2\ell+1} R^{2\ell+1}
\end{align}
or
\begin{align}
G\sigma_\ell & =
\dfrac{\ell-1}{4(\ell+2)}\dfrac{b_\ell}{(2\ell-1)!!}\left(\dfrac{GM}{c_0^2}\right)^{2\ell+1}\nonumber\\
            & =
\dfrac{\ell-1}{4(\ell+2)}\dfrac{j_\ell}{(2\ell-1)!!} R^{2\ell+1},
\end{align}
where we used the notation $j_\ell\equiv c^{2\l+1}b_\l$ 
of Eq.~\eqref{eq:cb}. As announced, we see that, modulo a numerical 
coefficient, the odd-parity analog of the 
$R$-scaled Love number $k_\ell$ is the combination
$c^{2\ell+1}b_\ell$, Eq.~\eqref{eq:cb}.
In the odd-parity, quadrupolar case ($\ell=2$) the above link reads
\be
\label{eq:b2_j2}
G\sigma_2 = \dfrac{1}{48} j_2 R^5 =
\dfrac{1}{48}b_2\left(\dfrac{GM}{c_0^2}\right)^5 , 
\ee
while the explicit expression of $j_2 = c^5 b_2$ reads
\begin{widetext}
\be
\label{eq:j2}
j_2 =\frac{96 c^5 (2 c-1) (y-3)}{5 \left(2 c \left(12 (y+1) c^4+2 (y-3) c^3+2 (y-3) c^2+3 (y-3) c-3 y+9\right)+3 (2 c-1) (y-3) \log (1-2 c)\right)}.
\ee
\end{widetext}

\section{Computation of the shape Love number $\boldsymbol{h_\ell}$ }
\label{sec:sec6}
We have indicated above (following a recent study of the gravitational
polarizability of black holes~\cite{DL09}) how one can generalize to a
relativistic context the shape tidal constant $h_\ell$ (the ``first Love
number'') introduced in Newtonian theory. Let us first point out that, though
in general there is no direct connection between $h_\ell$ and $k_\ell$, there
is a simple relation between them in the case where the deformed object is a
ball of (barotropic) perfect fluid, treated in Newtonian gravity. Indeed, in
the Newtonian theory of tidal deformation 
(see~\cite{Kopal,Mora:2003wt}), we have the result that an external
``disturbing'' tidal potential 
\be
U^{\rm disturb} = \sum_\ell c_\ell r^\ell P_\ell(\cos\theta)
\ee
deforms the constant-pressure (and constant-density) level surfaces ($p=p(a)$)
of a fluid star into
\be
r(a) = a\left(1+\sum_\ell f_\ell(a)P_\ell(\cos\theta)\right).
\ee
Here $f_\ell(a)=\left(\delta r/r\right)_\ell$ satisfies the Clairaut 
equation (in which $\bar{\rho}(a)$ indicates the mean density within 
$0\leq r\leq a$)
\be
\label{clairaut}
a^2 f''_\ell + \dfrac{6\rho(a)}{\bar{\rho}(a)}\left(af'_\ell + f_\ell\right)-\ell(\ell+1)f_\ell=0,
\ee
and is related to the disturbing tidal coefficient $c_\ell$ via
\be
f_\ell(A) = \dfrac{(2\ell+1)c_\ell A^{\ell+1}}{G(\ell+\eta_\ell)M},
\ee
where $A$ is the surface value of $a$ (i.e. $A\simeq R$, the undisturbed radius)
and where $\eta_\ell = \left[af'_\ell(a)/f_\ell(a)\right]_{a=A}$ denotes 
the surface logarithmic derivative of $f_\ell(a)$.
The latter quantity is related to the ``second'' Love
number $k_\ell$ via
\be
\label{eq78}
k_\ell=\dfrac{\ell+1-\eta_\ell}{2(\ell+\eta_\ell)}.
\ee
On the other hand,
using the definition of $h_\ell$, i.e.
\be
\left(\dfrac{\delta R}{R}\right)_\ell = f_\ell(A) = h_\ell\dfrac{c_\ell R^\ell}{GM/R},
\ee
we find
\be
\label{eq80}
h_\ell = \dfrac{2\ell+1}{\ell+\eta_\ell}.
\ee
By eliminating $\eta_\ell$ between Eq.~\eqref{eq78}
and Eq.~\eqref{eq80}, we finally get a simple 
relation between $h_\ell$ and $k_\ell$, namely
\be
\label{link_h_k}
h_\ell = 1+2 k_\ell .
\ee
For instance, a Newtonian $\gamma=2$ polytrope has a density profile
$\rho(r)\propto \sin x/x$, where $x\equiv \pi r/R$, from which one
deduces, using either the Clairaut equation, or the Newtonian limit
of the ($\ell=2$) even-parity master 
equation~\eqref{eq:H}, namely~\cite{Hinderer:2007mb}
\be
H'' + \dfrac{2}{r}H' + \left(4\pi G\rho\dfrac{d\rho}{dp}-\dfrac{6}{r^2}\right)H=0,
\ee
that $H\propto x^{-1/2}J_{5/2}(x)$. [Note that this result is 
misprinted as  $x^{+1/2}J_{5/2}(x)$ in~\cite{Hinderer:2007mb}].
This leads to~\cite{Hinderer:2007mb}
\be
k_2^N(\gamma=2)=-\dfrac{1}{2} + \dfrac{15}{2\pi^2}\simeq 0.25991,
\ee
and therefore
\be
h_2^N = \dfrac{15}{\pi^2}\simeq 1.51982 .
\ee
When generalizing the definition of $h_\ell$ to the relativistic context it
seems that we lose the existence of a functional relation between $h_\ell$
and $k_\ell$. Let us explicate the meaning and implementation of the
relativistic version of $h_\ell$, Eq.~\eqref{eq:def_hl}. First, we define
$\delta R/R$ as the fractional deformation of a sphere, embedded in an
auxiliary 3-dimensional Euclidean manifold, such that the inner geometry of
this deformed, embedded sphere is equal to the inner geometry (induced by the
ambient curved spacetime) of the real, tidally-deformed neutron star surface
(considered at fixed coordinate time). In general, one would need to consider
the Gaussian curvatures of both surfaces to express their identity (as done in
the black hole case~\cite{DL09}). Here, things are simpler because we are
using the Regge-Wheeler gauge. In that gauge, it is easily seen that the inner
metric of the surface $r=r(\theta,\varphi)$ of an (even-parity) tidally
deformed star is
\begin{align}
ds^2 &=
\left(r(\theta,\varphi)\right)^2\left(1-K\right)d\Omega^2\nonumber\\
&=R_0^2\left[1+
\left(2\dfrac{\delta r}{r}-K\right)\right]d\Omega^2,
\end{align}
where $r(\theta,\varphi)=R_0\left(1 +\delta r/r\right)$
is the radial coordinate location of the star's surface.
[Here we absorbed the $Y_{\lm}(\theta,\varphi)$ factors in $K$ and $\delta r/r$.]
Because this inner
metric is {\it conformal} to the usual sphere $S^2$ of unit radius 
$d\Omega^2=d\theta^2 + \sin^2\theta d\varphi^2$, it is easily checked 
that it would be the inner
geometry of a flat-embedded sphere (linearly) deformed by $\delta R/R$, with
\be
\dfrac{\delta R}{R} = \dfrac{\delta r}{r}-\dfrac{1}{2}K .
\ee
We can further compute the value of the coordinate deformation $\delta r/r$ 
by using the fact that the logarithmic enthalpy 
$h=\int_0^p dp/(e+p)$ must vanish on the star surface. 
Since $h(r) =h_0(r) + \delta h(r)$, where $h_0(r)$ is the enthalpy 
of the background undeformed star, and 
where $\delta h=-H/2$, we then find
\be
\delta r =\dfrac{1}{2}\left(\dfrac{H}{h'}\right)_{r=R}=\dfrac{1}{2}\left(\dfrac{e+p}{p'}
  H\right)_{r=R},
\ee
where the prime denotes $d/dr$. Finally, we have the 
``flat-equivalent shape deformation'':
\be
\dfrac{\delta R}{R} = \dfrac{1}{2}\left(\dfrac{e+p}{rp'}H-K\right)_{r=R}.
\ee
As we said above, the metric variable $K$ can be expressed as a linear
combination~\cite{Lindblom:1997un} of $H$ and $H'$, namely $K=\alpha_1H'+\alpha_2 H$.
The coefficients $\alpha_1$ and $\alpha_2$ evaluated on the unperturbed
surface $r=R$ of the star read
\begin{align}
\alpha_1 &= \dfrac{2cR}{(\ell-1)(\ell+2)} ,\\
\alpha_2 &= \dfrac{1}{(\ell-1)(\ell+2)}\left\{\ell(\ell+1)+\dfrac{4 c^2}{1-2c}-2(1-2c)\right\}.
\end{align}
In addition, using the TOV equations, we also have on the star surface
\be
\dfrac{rp'}{e+p} = -\dfrac{c}{1-2c}.
\ee
By replacing (on the surface) also $RH'=y H(R)$, we finally obtain
\be
\dfrac{\delta R}{R}=-\dfrac{1}{2}H(R)\left\{\dfrac{1-2c}{c}+\dfrac{2c
  y}{(\ell-1)(\ell+2)}+\alpha_2\right\}.
\ee
At this stage, we have obtained an expression of the form $\delta
R/R=H(R)f(c,y)$. To proceed further and compute $h_\ell$, it remains 
to obtain the value of $U_\ell^{\rm disturb}$.
Following Ref.~\cite{DL09}, and the spirit of the Newtonian definition of
$h_\ell$, we define $-2U_\ell^{\rm disturb}(R)$ as being the analytic
continuation at radius $r=R$ of the leading asymptotically growing piece in
$H$, i.e. the part of $H^{\rm growing}=a_P \hat{P}_{\ell 2}$ which grows as
$r^\ell$. In other words, we define it by the Newtonian-looking formula
\be
U_\ell^{\rm disturb}(R) = -\dfrac{1}{2}a_P \left(\dfrac{R}{M}\right)^\ell .
\ee
We can then compute $U_\ell^{\rm disturb}(R)$ in terms of the full value of
$H$ on the surface $H(R)=a_P\hat{P}_{\ell2}(x) + a_Q\hat{Q}_{\ell2}(x)$
(with, we recall, $x=R/M-1=1/c-1$) by separating two ``correcting factors''
out of $H(R)$, namely
\be
-\dfrac{1}{2}H(R) = \left[c^\ell \hat{P}_{\ell2}(x)\right]\left[1+a_\ell\dfrac{\hat{Q}_{\ell2}(x)}
{\hat{P}_{\ell2}(x)}\right]U^{\rm disturb}_\l(R).
\ee
Putting all the pieces together, and inserting our general
result~\eqref{eq:a_ell} for $a_\ell=a_Q/a_P$, we obtain the following
final result for $h_\ell$:
\begin{widetext}
\begin{equation}
\label{eq:hl}
h_{\ell} = c^{\ell+1}\left.\hat{P}_{\ell 2}(x)
\left\{ \dfrac{1-2c}{c}+\dfrac{1}{(\ell-1)(\ell+2)} \left[2 c y_\ell
  +\ell(\ell+1)+\dfrac{4 c^2}{1-2c}-2(1-2c) \right] \right\}
\left(1-\dfrac{\de_x\log \hat{P}_{\ell2}(x)-c y_\ell}{\de_x\log
  \hat{Q}_{\ell2}(x)-c y_\ell}\right)\right|_{x=1/c-1}.
\end{equation}
\end{widetext}

\section{Results for the even-parity 
tidal coefficient $\boldsymbol{\mu}_\ell$.}
\label{sec:sec7}

Having explained how to compute the various tidal constants $\mu_\ell$ (or
$k_\ell$), $\sigma_\ell$ (or $j_\l$) and $h_\ell$, let us discuss the dependence of these
quantities on the compactness $c=GM/(c_0^2 R)$, for various kinds of EOS.
As we already mentioned, we shall consider a sample of EOS.

\subsection{Polytropic Equations of State}

First, we consider two kinds of relativistic polytropes:
 the ``energy-polytrope'', or
$e$-polytrope, such that $p=\kappa e^\gamma$, where $e$ is the total energy
density, and the ``rest-mass-polytrope'', or $\mu$-polytrope, with
$p=\kappa\mu^\gamma$ and $e=\mu + p/(\gamma-1)$, where $\mu=n m_b$ is the baryon
rest-mass density.
For these polytropes, we shall focus on the adiabatic index $\gamma=2$, which
is known to give a rather good representation of the overall
characteristics of neutron stars. We shall also briefly explore what happens
when $\gamma$ takes values larger (or smaller) than $2$. 
In particular, we shall discuss, in the next subsection, the
limit $\gamma\to\infty$, which leads to an {\it incompressible} model, with
uniform energy density $e$.  Let us note that
the compactness of the $\gamma=2$ $e$-polytrope models ranges between $0$ (for
a formally vanishingly small central pressure) and 0.265 for the maximum mass
model, while the compactness of the $\mu$-polytrope ranges between $0$ and $0.2145$.
Note that the limit of a vanishing compactness $c=GM/(c_0^2R)\to 0$ formally corresponds
to the Newtonian limit. Augmenting $\gamma$ allows one to reach higher
compactnesses, and thereby to better explore the effects of general
relativistic strong-field gravity. 
In particular, the incompressible limit, 
$\gamma\to\infty$, yields a range of compactnesses which extends up to
$c_{\rm max}=4/9=0.4444\dots$, quite close to the ``black-hole compactness''
$c^{\rm BH}=1/2=0.5$. We note that a theorem guarantees that $4/9$ is the
highest possible compactness of a general relativistic perfect fluid ball
(see e.g.~\cite{wald,straumann}).

\begin{figure}[t]
\begin{center}
\includegraphics[width=80 mm, height=65mm]{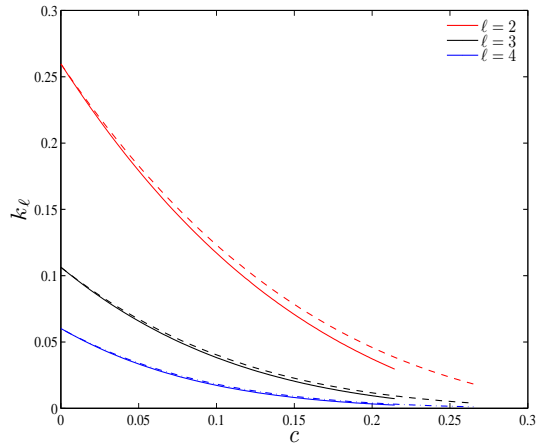}
\caption{\label{fig:fig1}Polytropic EOS: gravito-electric Love numbers $k_\ell$ 
(or apsidal constants) for $\ell=2,3,4$ versus compactness $c=M/R$.
We use two different polytropic EOS's, either of the rest-mass type
($p=\kappa\mu^\gamma$; solid lines) or of the energy type
($p=\kappa e^\gamma$; dashed lines).
For both EOS's we use $\gamma=2$. Note 
that the maximum compactness allowed by the $e$-polytrope is 
larger than that for the $\mu$-polytrope.}
  \end{center}
\end{figure}

Figure~\ref{fig:fig1} exhibits the dependence of the dimensionless,
even-parity, Love number $k_\ell$ on the compactness $c$ of $\gamma=2$
polytropes for three values of the multipole order: $\ell=2,3$ and $4$. 
The results for the $\mu$-polytrope (solid lines) are compared with those 
for the $e$-polytrope (dashed line). The limiting values of $k_\ell$
for $c\to 0$ (which numerically means $c\simeq 10^{-4}$)
do agree, as they should, with the known Newtonian
results~\cite{BO54,Berti:2007cd}.
In particular, $k_2^N(\gamma=2) = 0.25991$ (as mentioned above)
$k_3^N(\gamma=2) = 0.10645 $, and $k_4^N(\gamma=2)= 0.06024$.

The most striking structure of Fig.~\ref{fig:fig1} is the very strong decrease
of $k_{\ell}$ with increasing compactness. For typical neutron star
compactness, say $c\sim 0.15$, the general relativistic value of $k_\ell$ is
about $4$ times smaller than its Newtonian estimate $k_\ell^N$. This might
have an important negative impact on the measurability of neutron star
characteristics through gravitational-wave observations. Leaving this issue to
a future investigation~\cite{DNT}, let us focus here on a deeper understanding
of the $c$-sensitivity of $k_\ell$.

The main origin of the strong decrease of $k_\ell$, when $c$ increases, is the
universal presence of an overall factor $(1-2c)^2$ in $k_\ell$. As discussed
above, the presence of this term is linked to the no-hair property of black
holes, i.e. the fact that the star-rooted contribution $\propto a_Q
\hat{Q}_{\ell2}$ in the metric variable $H$ around a tidally-deformed star
becomes singular, in the black hole limit $R\to 2GM/c_0^2$.
In addition, there are other $c$-dependent effects that tend to decrease the
value of $k_\ell$. This is illustrated in Fig.~\ref{fig:fig2}
which plots the ratios $\hat{k}_\ell \equiv k_\ell/[k_\ell^{\rm N}(1-2c)^2]$ for
the $\gamma=2$, $\mu$-polytropic EOS.
In the case of $k_2$, Fig.~\ref{fig:fig2} shows that the ``normalized'' Love
number $\hat{k}_2$ is, to a good approximation, a linearly decreasing function 
of $c$, $\hat{k}_2(c)\simeq 1-\beta c$, with a slope $\beta\sim 3$. In other
words, the $c$-dependence of $k_2$ (for $\gamma=2$) is approximately 
describable as
\be
k_2(c)\simeq k_2^N(1-2c)^2(1-\beta c)\qquad\quad (\gamma=2),
\ee
with $\beta\simeq 3$.  To get a more accurate
representation, one must include more terms in the $c$-expansion of the 
``normalized'' $k_2$, or more generally $k_\ell$, say
\begin{equation}
\label{eq:fit}
k_{\ell} = k_{\ell}^N (1- 2c)^2\sum_{n=0}^4 a_n^\ell c^n.
\end{equation}
\begin{figure}[t]
\begin{center}
\includegraphics[width=80 mm, height=65mm]{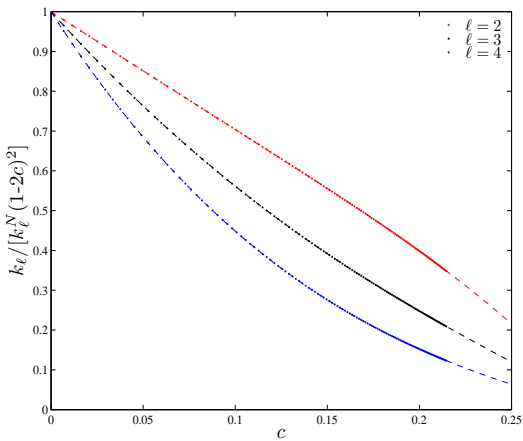}
\caption{\label{fig:fig2} Normalized Love numbers versus $c$
for a $\gamma=2$ $\mu$-polytrope (points); and
 performance of the fitting with the
template given by Eq.~\eqref{eq:fit} (dashed lines). The coefficients of the 
fit for each $\ell$ are listed in Table~\ref{tab:table1}.}
  \end{center}
\end{figure}
Such a nonlinear fit yields an extremely accurate representation of the
$c$-dependence of $k_\ell(c)$. The performance of such fits is illustrated in
Fig.~\ref{fig:fig2} (dashed lines) and the best fit values of the coefficients
$a_n^\ell$, $0\leq n\leq 4$ (fitted over the full range $0<c<c_{\rm max}$)
are listed in Table~\ref{tab:table1}.
Note that, if we were to trust these fits beyond the range $0<c<c_{\rm
  max}(\gamma)$ where $k_\ell(c)$ is defined, they would predict that 
$k_\ell(c)$ vanishes for a value $c_*^\ell$ slightly smaller than 
$c^{\rm BH}=1/2$, and would become negative before vanishing 
again (now quadratically) at $c=c^{\rm  BH}$. The critical 
value $c_*^\ell$ is around  $1/3$, 
and approximately independent of $\ell$.
\begin{table}[t]
\caption{\label{tab:table1}Fitting coefficients for $k_\ell$ as defined in
 Eq.~\eqref{eq:fit} for a $\gamma=2$ $\mu$-polytrope,  up to $\ell=4$. }
\begin{center}
  \begin{ruledtabular}
  \begin{tabular}{cccc}
    $\ell$   & $2 $ 
               & $3$
               & $4$ \\
    \hline \hline
 $a_0^{\ell}$    &     0.9991 &   0.9997   &  0.9998     \\
 $a_1^{\ell}$    &    -2.9287 &  -5.0933   & -7.1938     \\  
 $a_2^{\ell}$    &    -1.1373 &   7.2008   &  18.9509    \\
 $a_3^{\ell}$    &    14.0013 &   1.0826   & -21.8488    \\
 $a_4^{\ell}$    &   -50.9711 &  -18.7750  &  4.9031     \\
  \end{tabular}
\end{ruledtabular}
\end{center}
\end{table}%

\subsection{Incompressible Equation of State}
\label{sbsc:incompressible}

To further explore what happens for large compactnesses, we have studied in
detail the limit $\gamma\to\infty$, i.e. the incompressible EOS, $e=const$.
Let us recall  that in this case the
TOV equations can be solved analytically giving
\begin{equation}
p = e\dfrac{\sqrt{1-2c}-\sqrt{1-2cr^2}}{\sqrt{1-2c r^2} - 3 \sqrt{1-2c}}.
\end{equation}
Here $r$ denotes the dimensionless ratio $r^{\rm phys}/R$, so that $0\leq r\leq 1$.
Note that the central values ($r=0$) of the pressure are
$p_c = e(\sqrt{1-2c}-1)/\left(1-3\sqrt{1-2c}\right)$, so that $p_c\to\infty$ when
$c\to 4/9$, which shows that $c_{\rm max}=4/9$ is the maximum compactness
reachable by an incompressible star.
\begin{figure}[t]
\begin{center}
\includegraphics[width=80 mm, height=65mm]{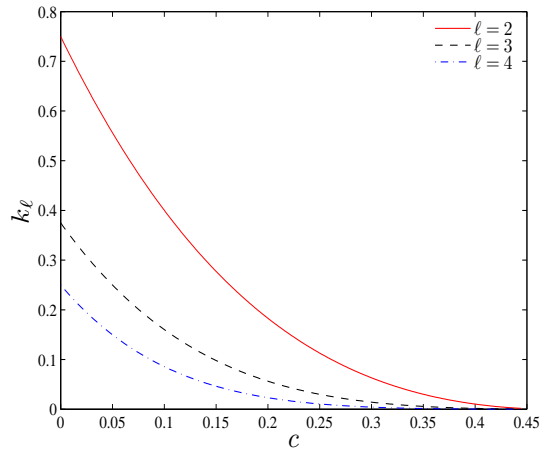}
\caption{\label{fig:fig3} Incompressible EOS: gravito-electric Love numbers
  $k_\ell$ (or apsidal constants) for $\ell=2,3,4$ versus compactness $c=M/R$.}
\vspace{-5mm}
  \end{center}
\end{figure}

Let us now discuss the computation of $k_\ell$ for an incompressible star. The
limit $e=\text{const}$ creates a technical problem in the use of the master
equation~\eqref{eq:H}. Indeed, the coefficient $C_0$ of $H$, Eq.~\eqref{eq:C0},
contains a contribution $\propto (e+p)de/dp$ which formally vanishes in the
incompressible limit $e=\text{const}$. However, as $e(r)$ is an inverted
step function which vanishes outside the star, 
$e(r) = e_0\left(1-\theta(r-1)\right)$, this term actually contributes 
a term $\propto \delta(r-1)$ which must crucially be taken into account 
in the computation of $k_\ell$. We can then proceed as follows.
First, one numerically integrates the incompressible limit of Eq.~\eqref{eq:H}
in the open interval $0<r<1$ representing the interior of the star. The output
of this integration is the value of the logarithmic derivative of $H$ at
$r=1^-$, say $y_{in}\equiv y(R^-)$. Second, one corrects this value into the
value $y_{out}\equiv y(R^+)$ just outside the star. To compute the correction
let us evaluate the ``strength'' of the delta-function singularity in the only
singular piece of $C_0$, namely $C_0^{\rm sing}=4\pi G e^\lambda (e+p)de/dp$.
Using the TOV equations giving the radial derivative of $p$, we can write
(reverting to general units, with $r=r^{\rm phys}$)
\be
C_0^{\rm sing}=-\dfrac{4\pi G r^2}{m(r) + 4\pi G r^3 p}\dfrac{de}{dr}.
\ee
In the incompressible limit $de/dr = -e_0\delta(r-R)$, $m(r) = (4\pi G/3)e_0
R^3$ and $p(R)=0$, so that
\be
C_0^{\rm sing} = + \, \dfrac{3}{R}\delta(r-R).
\ee
Then the effect of the singular term in Eq.~\eqref{eq:H}, or, more clearly,
in the corresponding Riccati equation for $y(r)=rH'/H$,
\be
ry' + y(y-1) + r C_1 y + r^2 C_0 = 0 ,
\ee
is easily found to introduce a step function singularity in $y(r)$ with 
strength $y^{\rm sing}(r) = -3\theta(r-R)$.
This shows that the correct  value $y_{\rm out}=y(R^+)$ to be used in
evaluating $k_\ell$ is (independently of the value of $\ell$)
\be
\label{minus_3}
y^{\rm out}_{\ell} = y^{\rm in}_\ell - 3.
\ee
As a check on this result, we can consider the Newtonian limit of
Eq.~\eqref{eq:H}.
In this limit, the exterior solutions $\hat{P}_{\ell2}(x)$ and $\hat{Q}_{\ell2}(x)$
reduce to $(r/M)^\ell$ and $(M/r)^{\ell+1}$ respectively, so that one has
\be
k_\ell^N = \dfrac{1}{2}\dfrac{\ell-y}{\ell+1+y},
\ee
which generalizes the $\ell=2$ result of~\cite{Hinderer:2007mb} 
to an arbitrary $\ell$.
Then, the incompressible limit, in the interior, of the Newtonian limit
of Eq.~\eqref{eq:H} reads
\be
H'' + \dfrac{2}{r}H' - \dfrac{\ell(\ell+1)}{r^2}H = 0 ,
\ee
which coincides with the exterior, Newtonian equation for $H$, with general
solution $H=a_P (r/M)^\ell + a_Q (M/r)^{\ell+1}$. Regularity at the origin
selects the $a_P$ term, so that 
$y_\ell^{\rm in}(r) = \ell = y^{\rm  in}_\ell(R)$. Then, Eq.~\eqref{minus_3}
determines
\be
y_\ell^{{\rm out}\,N} = \ell-3 \ ,
\ee
so that
\be
\label{inc:newton}
k_\ell^{N\,\text{(incomp)}} = \dfrac{3}{4(\ell-1)}.
\ee
This result agrees with the known result for a limiting $\gamma=\infty$
($n=0$) polytrope~\cite{BO54}.

Figure~\ref{fig:fig3} shows our numerical results for the $k_\ell$ Love
numbers of a {\it relativistic} incompressible star, as function of $c$, 
in the range $0\leq c \leq c_{\rm max}=4/9$. We exhibit
the three first multipolar order $\ell=2,3$ and $4$. Note that the $c\to 0$
values of $k_\ell(c)$ agree with the Newtonian limit, Eq.~\eqref{inc:newton}.
\begin{figure*}[t]
\begin{center}
\includegraphics[width=85 mm, height=70mm]{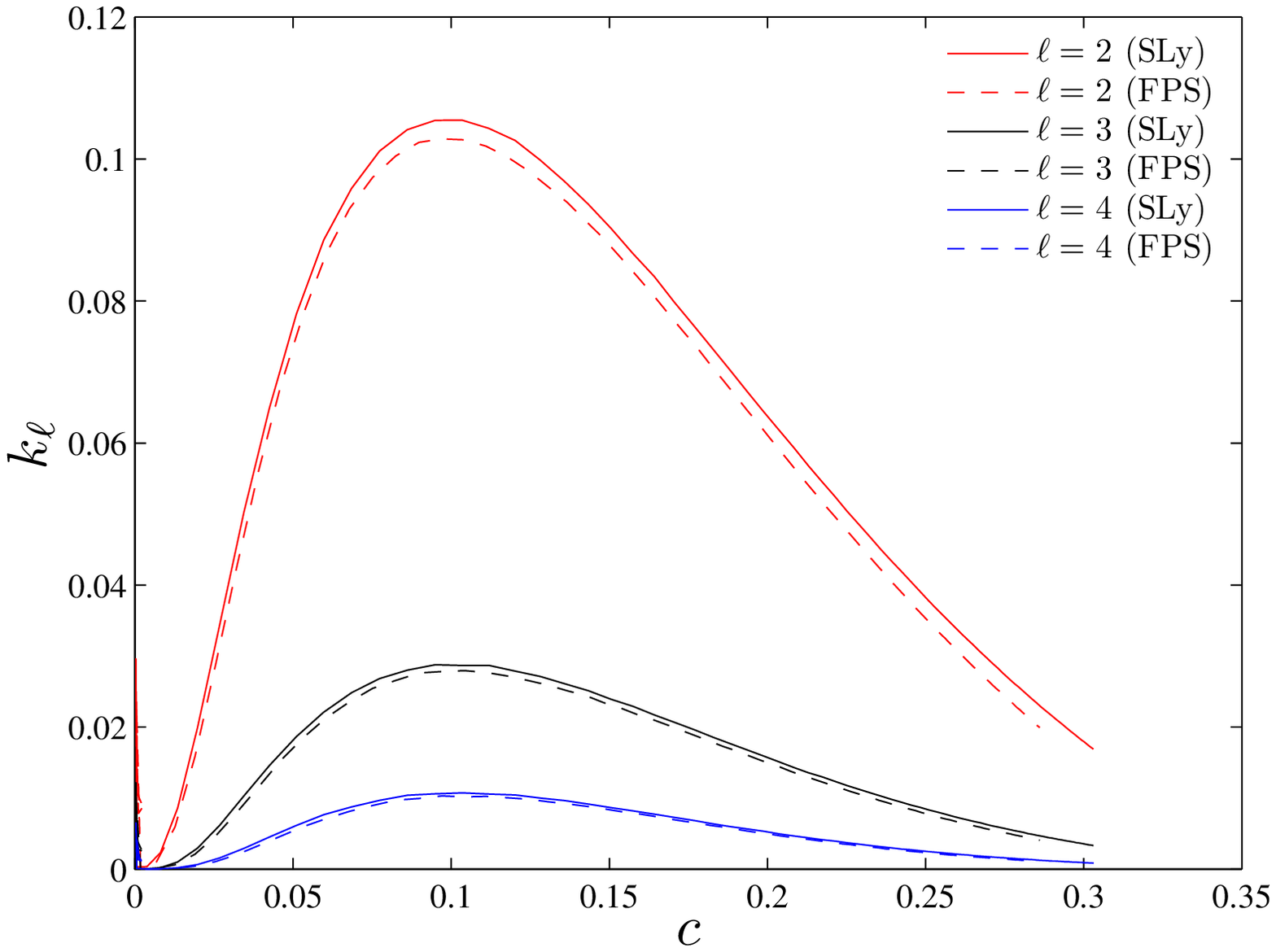}
\includegraphics[width=85 mm, height=70mm]{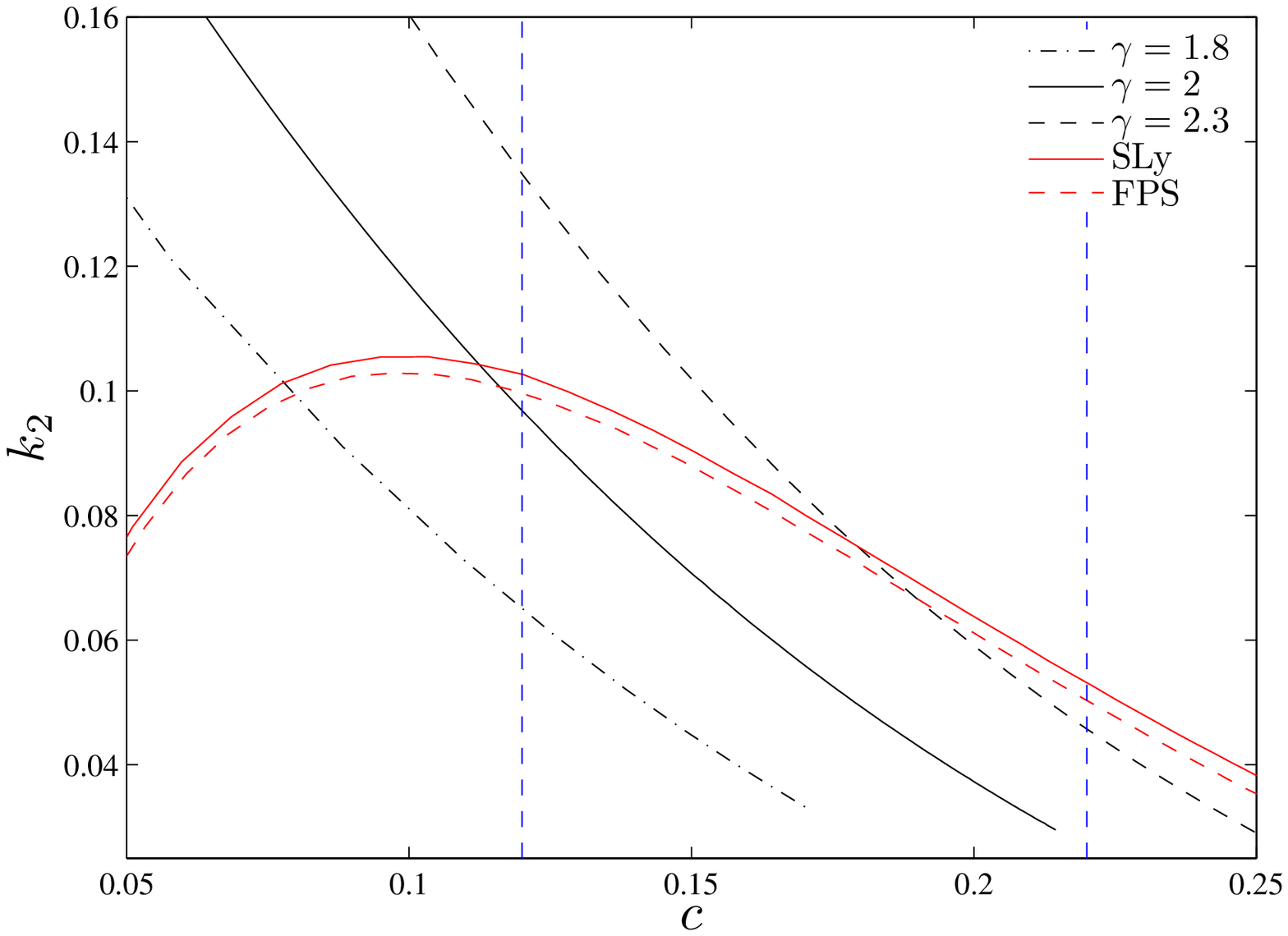}
\caption{\label{fig:fig4}The gravito-electric Love numbers $k_\ell$ (or
  apsidal constants) for $\ell=2,3,4$ versus compactness $c=M/R$ for the
  two tabulated ``realistic'' equation of state FPS and SLy (left panel). 
  Right panel: comparison between $k_2$ from various 
  relativistic $\mu$-polytropes (with different $\gamma$) and the 
   FPS and SLy realistic EOS's.}
  \end{center}
\end{figure*}
The phenomenon of the ``quenching'' of $k_\ell$ as $c$ increases is even more
striking, in this incompressible case, than in the $\gamma=2$ case considered
in Fig.~\ref{fig:fig1} above. Note, in particular, the very small values
reached by $k_\ell$ for the maximum compactness $c=4/9$. 
To understand better the ``quenching'' of $k_\ell$ by strong-field effects, we
have analytically studied the incompressible model in the limit of maximum
compactness $c\to 4/9$. This limit is singular (because $p_c\to\infty$), but
is amenable to a full analytical treatment of $k_\ell$. This analytical study
is thereby a useful {\it strong-field} analog of the analytical study of the
$\gamma=2$ model in the {\it weak-field} (Newtonian) limit. Let us sketch how
one can analytically solve the $\gamma=\infty$, $c\to 4/9$ model. 
First,  by introducing the variable
\begin{equation}
\label{eq:x}
x = \sqrt{1-\dfrac{8}{9} r^2},
\end{equation}
Equation~\eqref{eq:H} becomes
\begin{align}
(x^2 -1)^2\dfrac{d^2H}{d x^2} &+ \left(4 x^3 + x^2 -4 x
 -1\right)\dfrac{dH}{dx}\nonumber\\
& + \left(2 x^2 + x - \ell(\ell+1)-1\right)H = 0.
\end{align}
We found two exact, analytical solutions of this equation, which are both of
the form $(1-x)^\alpha(1+x)^\beta$ with some rational exponents $\alpha$,
$\beta$. More precisely, either
\be
\alpha = \dfrac{\ell-1}{2},\qquad \beta=-\dfrac{\ell+1}{2},
\ee
or 
\be
\alpha = -\dfrac{\ell+2}{2},\qquad \beta= \dfrac{\ell}{2}.
\ee
Regularity at the origin~\footnote{Note, however, that because of the singular
  behavior of $p(r)$ at the origin, the ``regular'' solution $H(r)$ is less
  regular than usual: $H(r)\propto r^{\ell-1}$ instead of $r^\ell$.}
selects the first solution, namely
\begin{equation}
H(x) = (1-x)^{(\ell-1)/2}(1+x)^{-(\ell+1)/2}.
\end{equation}
As a result, the {\it interior} value of the logarithmic 
derivative $y_{\rm int}$ of $H$ has the simple form
\begin{equation}
y(x) = \dfrac{\ell}{x}-1 .
\end{equation}
Adding the effect of the $\delta$-function at the 
surface ( $r=1$, i.e. $x=1/3$), Eq.~\eqref{minus_3}
finally leads to
\be
y_\ell^{\rm out} = y_{\rm int}(1/3)-3 = 3\ell - 4.
\ee
Inserting this result in our general result~\eqref{eq:kl}
for $k_\ell$ gives an analytical expression for $k_\ell^{(\rm incomp)}(c_{\rm max})$.
For instance, one finds the following value for $\ell=2$
\be
k_2^{(\rm incomp)}(c_{\rm max})=\dfrac{4096}{10935(308-81\log 3)}\simeq 0.0017103 .
\ee
Note the striking quenching of $k_2$, by nearly 3 orders of magnitude,
from the $c\to 0$ value $k_2^N=0.75$ to this result for $c_{\rm max}=4/9$.

\subsection{Realistic Equations of State}
\label{sec:realistic}

Before discussing our results for the other tidal coefficients ($b_\ell$,
$\sigma_\ell$, $h_\ell$), let us end this section devoted to the $k_\ell$ (and
$\mu_\ell$) Love numbers by briefly considering the $k_\ell$, 
for the dominant quadrupolar order $\ell=2$, predicted by the two 
realistic (tabulated) EOS FPS and 
SLy\footnote{Since these EOS are given through tables, to use 
them in a numerical context it is necessary to interpolate between
the tabulated values. As in~\cite{Bernuzzi:2008fu} we use simple 
linear interpolation (instead of third-order Hermite or spline ones) 
to avoid the introduction of spurious oscillations in the speed of 
sound. See Ref.~\cite{Bernuzzi:2008fu} for further details.}.
We recall that we have chosen them here because they have
been used in some recent numerical relativity simulations
of coalescing neutron star binaries~\cite{Shibata:2005ss,Shibata:2005xz}.
 The maximum compactness of the two ``realistic'' EOS that we retained are 
$c^{\rm FPS}_{\rm max} = 0.2856 $ and $c_{\rm max}^{\rm SLy}=0.303$.

The corresponding results for $k_\ell$ are shown in the left 
panel of Fig.~\ref{fig:fig4} for $2\leq \ell\leq 4$. 
The right panel focuses only on the 
results for the $\ell=2$ case, that are plotted together with
several illustrative $\mu$-polytropes , namely $\gamma=1.8$, $2$ 
and $2.3$. The $\gamma=1.8$ polytrope illustrates the reason why the
``realistic'' EOS lead to a decrease of $k_\ell$ as $c\to 0$. 
This is Bacause of the  fact that the ``local'' adiabatic index
\be
\Gamma = \left(1+\dfrac{p}{e}\right)\dfrac{d\log p}{d\log e} = \dfrac{d\log p}{d\log\mu}
\ee
of these EOS varies with the density (or pressure). As shown, e.g., 
in the bottom panel of Fig.1 of Ref.~\cite{Bernuzzi:2008fu} 
for $\sim$ nuclear densities it stays in the 
range $2\lesssim \Gamma\lesssim 2.3$, while,
for lower densities (around neutron drip) it drops to low 
values  $\Gamma\lesssim 1$, before rising again towards
$\Gamma\sim 4/3$ for low densities. Let us recall in this
respect that Newtonian polytropes have a finite radius only for $\gamma>1.2$
($n<5$). When $\gamma\to1.2$, a Newtonian polytrope has finite mass, but its
radius $R$ tends to infinity. As $k_2$ uses a scaling of $\mu_2$ by a power of
$R$, this causes $k_2^N=k_2(c=0)$ to tend to zero as $\gamma\to 1.2$ (see~\cite{BO54}).
Anyway, the decrease of $k_2^{\rm realistic}(c)$ as $c\to 0$, linked to the
small value of $\Gamma$ for low (central) densities and pressures 
is a mathematical property which is physically irrelevant for
our main concern, namely the tidal properties of neutron stars. Indeed,
neutron stars have a minimal mass determined by setting the mean value of
$\Gamma$ equal to the critical value $\sim 4/3$ for radial stability
against collapse~\cite{Shapiro:1983du}. Moreover, we are mainly interested in
neutron star masses $\sim 1.4M_{\odot}$. Such neutron stars are expected to
have radii varying at most in the range $10\text{km}\lesssim R\lesssim
15\text{km}$, corresponding to compactnesses $0.13\lesssim c\lesssim 0.2$.
To be on the safe side, we shall consider the interval $0.12\leq c\leq 0.22$, 
which is indicated by vertical lines in the right panel of
Fig.~\ref{fig:fig4}. 
Focusing our attention on this interval, we can draw the following conclusions 
from the inspection of the right panel of Fig.~\ref{fig:fig4}:
\begin{enumerate}
\item The $\mu$-polytropes $\gamma=2$ and $\gamma=2.3$ approximately bracket 
      the $k_2(c)$ sequence predicted by the two realistic EOS's.
\item Actually, the two realistic EOS's that we retained here, FPS and SLy, 
      lead to rather close predictions for $k_2(c)$.
\item In the range $0.12\leq c\leq 0.22$ the two, ``realistic'' $k_2(c)$ can
      be approximately represented by the following linear fit 
\be
\label{k2_real_fit}
k_{2}^{(\text{FPS;\,SLy})}\simeq A-Bc,
\ee
\end{enumerate}
with $A \simeq 0.165  $ and $B \simeq 0.515 $.

\section{Results for the odd-parity tidal 
coefficients $\boldsymbol{\sigma_\ell}$}
\label{sec:sec8}

As explained above, the odd-parity tidal coefficients $\sigma_\ell$ is
obtained by solving the master equation~\eqref{eq:odd_master}. Let us start by
noticing that the formal Newtonian limit of this master equation is simply
\be
\label{psi_newt}
r^2\psi'' = \ell(\ell+1)\psi .
\ee
Indeed, all the matter-dependent contributions to Eq.~\eqref{eq:odd_master} are,
fractionally, of order $Gm(r)/(c_0^2 r)\sim 4\pi G r^2 e/c_0^2$
or $4\pi G r^2 p/c_0^2$, and vanish in the Newtonian limit $c\to 0$.
In this limit, Eq.~\eqref{psi_newt} does not contain any effect of the star,
and, in particular, is the same in the interior or in the exterior of the
star. This shows that the origin-regular solution of Eq.~\eqref{psi_newt} is,
everywhere, of the form
\be
\psi^N (r)\propto r^{\ell+1}
\ee 
and does not contain any ``decreasing'', $Q$-type contribution
$b_Q\psi_Q^N\propto r^{-\l}$.
This proves that the odd-parity Love number $b_\ell=b_Q/b_P$ {\it vanishes} in
the Newtonian limit. More precisely, as the first post-Newtonian corrections to
Eq.~\eqref{psi_newt} are fractionally of order $c=GM/(c_0^3 R)$, we can then
easily see that the $R$-normalized odd-parity Love number $j_\l=c^{2\ell+1}b_\ell$
will be of order $c$ as $c\to 0$ (in agreement with the
 results of \cite{Favata:2005da} concerning the $\l=2$ case).

As we have pointed out above that $b_\ell$ also contains a factor $1-2c$, we
conclude that $j_\ell$ vanishes {\it both} when $c\to 0$ and
$c\to 1/2$, and should qualitatively be of the type
\be
\label{eq:B}
j_\ell=c^{2\ell+1} b_\ell\simeq B_\l c (1-2c).
\ee
This approximate result suggests that a 1PN-accurate calculation of the
solution of the master equation~\eqref{eq:odd_master} should give us access to the
coefficient $B_\l$, and thereby, in view of Eq.~\eqref{eq:B}, to a global
understanding of the $c$-dependence of $j_\ell$. In particular, one expects
from~\eqref{eq:B} that $|j_\ell| = |c^{2\ell+1}b_\ell|$ will attain a maximum value
somewhere around $c=1/4$, with the value 
\be
|j_\l|_{\rm max}\sim \dfrac{|B_\l|}{8}.
\ee
\begin{figure}[t]
\begin{center}
\includegraphics[width=80 mm, height=65mm]{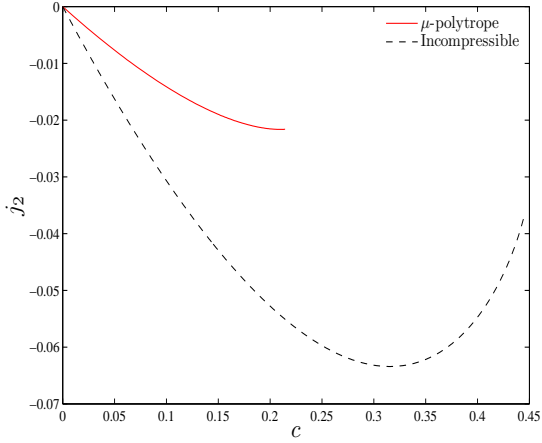}
\caption{\label{fig:fig5}The $j_2$, odd-parity, Love number for the
$\g=2$,  $\mu$-polytrope and for the incompressible EOS.}
  \end{center}
\end{figure}
Therefore, a 1PN computation of the coefficient $B$ gives an indication of
the maximum strength of the odd-parity Love number. We have analytically
computed the coefficient $B_\l$ (defined by $j_\ell=B_\l c+\O(c^2)$) by solving
Eq.~\eqref{eq:odd_master} by perturbation theory $\psi=\psi_0+\psi_1$. Here
$\psi_0=r^{\ell+1}$ is the solution of the $c\to 0$ limit,
Eq.~\eqref{psi_newt}, of Eq.~\eqref{eq:odd_master}, and $\psi_1$ is the first-order
effect of the matter terms in Eq.~\eqref{eq:odd_master}. Actually, we found
convenient to get $\psi=\psi_0+\psi_1$ by using Lagrange's method of
variation of constants: $\psi(r) = c_1(r)r^{\ell+1}+c_2(r) r^{-\ell}$, with
$c_1(r)\to 1$ and $c_2(r)\to 0$ as $r\to 0$. Skipping technical details, we
found that the logarithmic derivative 
\be
y=\dfrac{r\psi'}{\psi}\simeq (\ell+1)\left[1-\dfrac{2\l+1}{\l+1}\dfrac{c_2(r)}{c_1(r)}r^{-(2\l+1)}\right]
\ee
takes the following value at the star surface
\begin{align}
\label{yR}
y(R) = (\l+1)&\bigg\{1+\dfrac{\l+2}{\l+1}\dfrac{1}{R^{2\l+1}}\nonumber\\
&\times \int_0^Rdr r^{2\l}\left[2(\l-2)\dfrac{m}{r}+4\pi r^2(e-p)\right]\bigg\}.
\end{align}
The integral term in this result represents the 1PN correction Bacause of the 
presence of matter. On the other hand, the small-$c$ limit of the general
result~\eqref{eq:cb} yields for $\ell=2$ (chosen for simplicity) 
\be
j_2\simeq -\dfrac{y-3}{y+2}\simeq -\dfrac{y-3}{5} ,
\ee
where we used the fact $y=3+\O(c)$. Combining this result 
with Eq.~\eqref{yR} above yields 
\be
j_2 = c^5 b_2\simeq -\dfrac{4}{5R^5}\int_0^Rdr 4\pi G (e-p)r^6.
\ee
One can analytically compute this integral in the case of a
Newtonian polytrope with $\gamma=2$. Recalling that in this case
we have $e-p\simeq \rho c_0^2\simeq \rho_c c_0^2\sin x/x$,
with $x=\pi r/R$, the result is
\be
j_2 = B_2 c + \O(c^2)\qquad (\gamma=2)
\ee
with
\be
B_2 = -\dfrac{4}{5}\left(1-\dfrac{20}{\pi^2}+\dfrac{120}{\pi^4}\right)\simeq - 0.164395,
\ee
in agreement with \cite{Favata:2005da} 
(modulo normalization issues that we did not check).

In view of the reasoning above, we then expect that $j_2\simeq B_2 c(1-2c)$ 
will be {\it negative}, will vanish at $c=0$ and (formally) at $c=1/2$, and will
reach a minimum value 
$\left(j_2\right)^{\rm min}_{\gamma=2}\simeq B_2/8 =-0.02055$ around $c=1/4$.
For completeness, let us also mention that the case of an incompressible
neutron star leads to 
\be
B_2^{(\rm incomp)} = -\dfrac{12}{35}=-0.342857
\ee
and $\left(j_2\right)^{\rm min}_{\rm incomp}\simeq-0.042857$.
\begin{figure}[t]
\begin{center}
\includegraphics[width=80 mm, height=65mm]{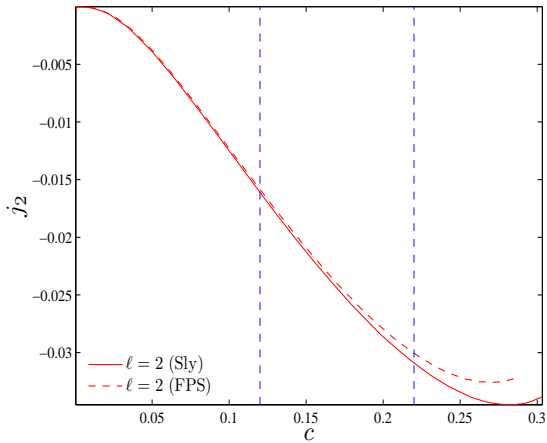}
\caption{\label{fig:fig5b}The $j_2$, odd-parity, Love number for
FPS and SLy EOS.}
  \end{center}
\end{figure}
We have qualitatively and semiquantitatively confirmed these results on the
$c$-dependence of the odd-parity Love number by numerically integrating
Eq.~\eqref{eq:odd_master}.
We display in Fig.~\ref{fig:fig5} the resulting odd-parity quadrupolar 
Love number $j_2$, versus $c$, for both a $\g=2$ $\mu$-polytrope 
(solid line) and an incompressible EOS (dashed line).
We have numerically checked that the slope at the origin of the $c$-axis is
indeed $B_2$ as analytically determined above. 
In both cases (though it is more evident in the incompressible case, where
higher values of $c$ are allowed) $j_2$ has a negative minimum before rising 
again towards zero. The numerically determined minimum of $j_2$ is $\text{min}(j_2)\simeq -0.0216$
(reached around $c\simeq 0.21$) for $\g=2$  and
$\text{min}(j_2)\simeq -0.0634$ (reached around $c\simeq 0.315$) 
for the incompressible EOS.

From the conceptual point of view, these results on the odd-parity Love number
(and, via Eq.~\eqref{eq:b2_j2}, on the corresponding magnetic-like tidal
coefficient $\sigma_\ell$) are interesting counterparts of the even-parity
results discussed above. They have points in common (their vanishing in the
formal limit $c\to1/2$), and they also strongly differ in other aspects:
$\sigma_\l$ vanishes when $c\to 0$, while $\mu_\l$ has a well-known Newtonian
limit; and $\sigma_\l$ is proportional to the first power of $1-2c$, while
$\mu_\l\propto (1-2c)^2$. Moreover, $\sigma_\l$, as naturally defined, is
negative, while $\mu_\l$ is positive. As, with our DSX-like normalization, the
interaction energies associated to both types of couplings are proportional
(modulo positive numerical constants) to $-M_L G_L = -\mu_\l G_L^2$ and
$-S_LH_L=-\sigma_\l H_L^2$ respectively, this sign difference can be
interpreted as being linked to the well-known, Lorentz-signature related, fact
that current-current interactions have always the opposite sign to
charge-charge, or mass-mass, interactions.
Concerning the formal vanishing of $\sigma_\ell$ as $c\to c^{\rm BH}=1/2$, the
same remarks we made above for the even-parity case apply here.
This fact is essentially, given the no-hair properties of black holes, 
a consistency check on the definition of $\sigma_\l$ as measuring a violation
of the ``effacing principle''. As said above, although it suggests that the
correct value of $\sigma_\l$ for black holes might be zero, it is far from
proving such a statement which has a meaning only within a more complex
nonlinear context.

From the practical point of view, an interesting output of the investigation
of $\sigma_2$ is that its numerical value happens to be quite small. Indeed,
for a $\g=2$ $\mu$-polytrope we have
\be
\dfrac{|G\sigma_2|}{R^5}^{\rm max}=\dfrac{1}{48}|j_2|^{\rm max}\simeq 4\times 10^{-4}.
\ee
We shall discuss in another work the precise dynamical meaning of this small
number (e.g. for the dynamics of binary neutron stars), but the appearance of
such a small number clearly means that it will be an enormous challenge to
measure it via gravitational-wave observations.

For completeness, we conclude this section by showing, in Fig.~\ref{fig:fig5b}, 
the behavior of $j_2$ also for the two different realistic EOS, FPS and SLy, 
that we have introduced above. 
Similarly to the case of $k_2$ (see Eq.~\eqref{k2_real_fit} above), 
in the range $0.12\leq c \leq 0.22$ 
(i.e., between the two dashed vertical lines) 
the two ``realistic'' $j_2(c)$ can be approximately represented by 
the following  linear fit
\be
j_2^{({\rm FPS};\;{\rm SLy})}\simeq A - B c,
\ee
with $A\simeq 1\times 10^{-4}$ and  $B\simeq 0.1411$.

\section{Results for the ``shape'' Love numbers $\boldsymbol{h_\ell}$}
\label{sec:sec9}
\begin{figure}[t]
\begin{center}
\includegraphics[width=80 mm, height=65mm]{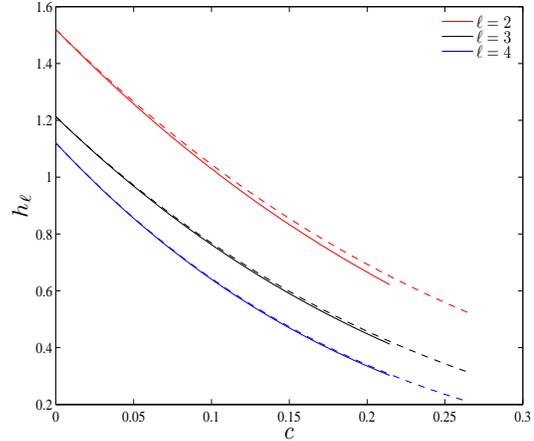}
\caption{\label{fig:fig6}Shape Love numbers $h_{\ell}$ versus $c$ for the 
two \hbox{$\g=2$} polytropic EOS: the $\mu$-polytrope (solid lines) and 
the  $e$-polytrope (dashed lines).}
\vspace{-5mm}
  \end{center}
\end{figure}
\begin{figure}[t]
\begin{center}
\includegraphics[width=80 mm, height=65mm]{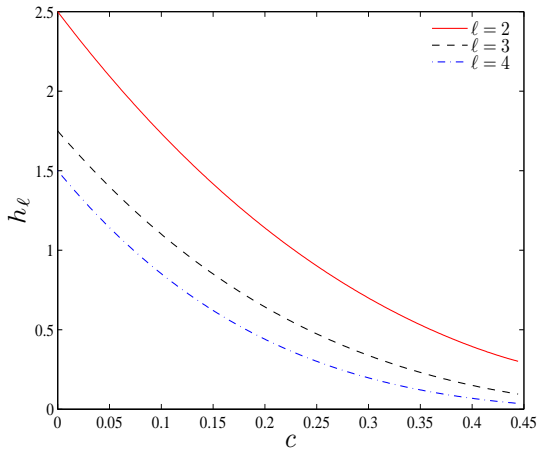}
\caption{\label{fig:fig7} Shape Love numbers $h_\ell$  versus $c$ for 
the incompressible model.}
  \end{center}
\end{figure}
Equation~\eqref{eq:hl} gave the final expression for the (even-parity) ``shape''
Love number $h_\l$. This expression contains several terms that are singular
as $c\to c^{\rm BH}=1/2$: (i) the long curly bracket contains an explicit term
$\propto 1/(1-2c)$; (ii) the logarithmic derivative of $\hat{Q}_{\l2}$
behaves, when $x\to 1$ (given that $\hat{Q}_{\l2}(x)\sim(x-1)^{-1}$), 
as $\de_x\log Q_{\l2}(x)\simeq - (x-1)^{-1}$; and (iii) the 
logarithmic derivative of $\hat{P}_{\ell2}$ behaves, when $x\to 1$ (given
that $\hat{P}_{\ell 2}(x)\simeq (x-1)^{+1}$, see below), as $\de_x\log\hat{P}_{\ell
  2}(x)\simeq +(x-1)^{-1}$.
The latter behaviors mean that, in the limit $R/(2M)\to 1$, 
the last factor in Eq.~\eqref{eq:hl} tends to $1+1=2$. As for the
$(1-2c)^{-1}$ singularity in the curly bracket, it is compensated by the
linear vanishing of $\hat{P}_{\ell2}(x)$ as $x\to 1$, indeed
\be
P_{\ell 2}(x)\sim (x^2-1)\dfrac{d^2 P_\l(x)}{dx^2}\sim x-1=\dfrac{1}{c}-2.
\ee
Finally, the formal ``black-hole limit'', $c\to c^{\rm BH}=1/2$ of the
``shape'' Love number $h_\l(c)$ is finite and nonzero. Actually, we found that
this limit agrees with the results of a recent direct investigation of the
``gravitational polarizability'' of a black hole~\cite{DL09}. The general
result for $h_\l^{\rm BH}$ can be found in the latter reference. Let us only
mention here the values of the first two ``shape'' Love numbers: 
\begin{align}
\lim_{c\to 1/2}h_2(c)&=h_2^{\rm BH} = \dfrac{1}{4} ,\\
\lim_{c\to 1/2}h_3(c)&=h_3^{\rm BH} = \dfrac{1}{20} .
\end{align}
Figure~\ref{fig:fig6} shows the results of inserting the numerically
determined value of $y_\ell(R)$ into the expression~\eqref{eq:hl} of
$h_\ell$. We give the results for the first three multipolar orders,
$\l=2,3,4$, and for the two $\gamma=2$ polytropes ($\mu$-polytrope and $e$-polytrope).
We have also investigated the results for the incompressible EOS
($\gamma\to\infty$). They are shown in Fig.~\ref{fig:fig7}.
This information is completed in Fig.~\ref{fig:fig8}, where we 
investigate the effect of changing the EOS on the $c$-behavior 
of the leading, quadrupolar, shape Love number $h_2$.
In all cases, we see that, somewhat similarly to the $k_2$ case, the strong
self-gravity of a neutron star tends to ``quench'' the value of $h_2$. For
instance, as we discussed above, the Newtonian limit of a $\g=2$ polytrope
yields $h_2^N(\gamma=2)=15/\pi^2\simeq 1.52$. As we see in Fig.~\ref{fig:fig8}
this value is reduced below $1$, i.e. by more than $33\%$, for typical neutron
star compactnesses. When exploring stronger self-gravity effects, notably for
the incompressible model, one gets an even more drastic quenching of $h_2$,
by an order of magnitude, from the Newtonian value
$h_2^{N\,(\text{incomp})}=2.5$ down to a value near 
the ``black hole'' value $h_2^{\rm BH}=1/4=0.25$.
\begin{figure}[t]
\begin{center}
\includegraphics[width=80 mm, height=65mm]{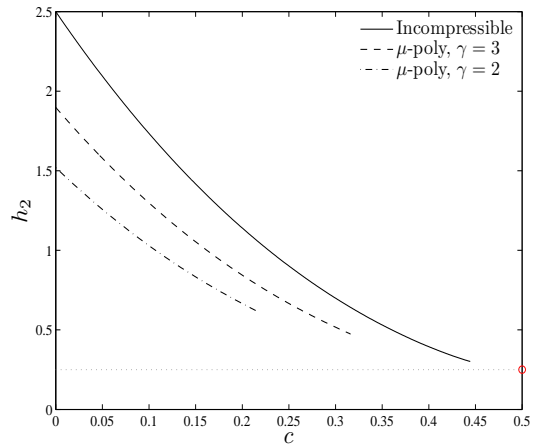}
\caption{\label{fig:fig8} Influence of the EOS on $h_2(c)$: 
comparison between the incompressible case and  $\mu$-polytropes 
with two values of the polytropic index: $\gamma=2$ and $\gamma=3$. 
The red circle on the right of the plot indicates the formal $c\to 1/2$
result.}
  \end{center}
\end{figure}
From the theoretical point of view, it is nice to see this continuity, as the
compactness increases, between the neutron-star case and the black-hole
case. Note that neither the no-hair property of black holes, nor the related
``effacing principle'', are relevant to the present result. What is relevant
is that the inner geometry of the horizon of a black hole is well defined and
that a black hole is an elastic object, like a neutron star.

Similarly to what we did for $k_{\ell}$ discussed above, it is also
convenient and useful to represent $h_{\ell}$ as a $c$-expansion
of the form
\begin{equation}
\label{eq:hfit}
h_\ell = h_\ell^N\sum_{n=0}^4 b^\ell_n c^n,
\end{equation}
where $h_\ell^N$ is the Newtonian value (obtained from $k_\ell^N$ 
through Eq.~\eqref{link_h_k}) and the coefficients $b_n^\ell$ are obtained 
from a fit. As an example, Table~\ref{tab:table2}
lists these coefficients for a $\gamma=2$, $\mu$-polytrope up to $\ell=4$
(i.e., they are obtained by fitting the solid lines in Fig.~\ref{fig:fig6}).

For completeness, we conclude this section by discussing the $h_\l$ results
for the two different realistic EOS, FPS and SLy, that we have introduced
above. In Fig.~\ref{fig:fig9} we display the $h_{\ell}$ Love numbers 
(for $\ell=2,3,4$) versus compactness $c$. The fact that 
$h_\l\to 1$ when $k_\l\to 0$ (because of the small value of the local
adiabatic index $\Gamma$ for low central densities and pressures) is
understood via the Newtonian link~\eqref{link_h_k}. 
\begin{figure}[t]
\begin{center}
\includegraphics[width=80 mm, height=65mm]{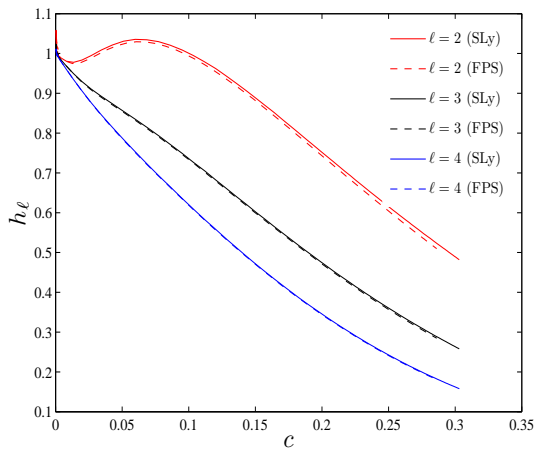}
\caption{\label{fig:fig9}Shape Love numbers $h_\ell$ versus $c$ for the two
  tabulated realistic Equation of State FPS and SLy.}
  \end{center}
\end{figure}
\begin{table}[t]
\caption{\label{tab:table2}Fitting coefficients for $h_\ell$ as defined in
 Eq.~\eqref{eq:hfit} for a $\gamma=2$ $\mu$-polytrope,  up to $\ell=4$. }
\begin{center}
  \begin{ruledtabular}
  \begin{tabular}{cccc}
    $\ell$   & $2 $ 
               & $3$
               & $4$ \\
    \hline \hline
 $b_0^{\ell}$    &     0.9999 &   0.9999   &  0.9999    \\
 $b_1^{\ell}$    &    -3.6764 &  -4.3700   & -5.2361    \\  
 $b_2^{\ell}$    &     4.5678 &   6.9775   &  10.4578   \\
 $b_3^{\ell}$    &    -0.0192 &  -4.1964   & -9.6026    \\
 $b_4^{\ell}$    &    -5.8466 &  -1.028    &  3.0415    \\ 
  \end{tabular}
\end{ruledtabular}
\end{center}
\end{table}%

\section{Conclusions}
\label{sec:conclusions}

We have studied the various tidal responses of neutron stars to external tidal
fields. We have considered both electric-type (even-parity) and magnetic-type
(odd-parity) external tidal fields. As indicated by Damour, Soffel and
Xu~\cite{Damour:1991yw} some time ago, one can correspondingly introduce two
types of linear response coefficients: an electric-type tidal coefficient
$G\mu_\l=[\text{length}]^{2\ell+1}$ measuring the $\l^{\rm th}$  mass
multipole $GM_L$ induced in a star by an external $\l^{\text{th}}$-order
(electric) tidal field $G_L$, and a magnetic-type tidal coefficient
$G\sigma_\l=[\text{length}]^{2\l+1}$ measuring the $\l^{\rm th}$ spin
multipole $GS_L$ induced in a star by an external $\l^{\text{th}}$-order
``magnetic'' tidal field $H_L$. Dividing $G\mu_\l$ and $G\sigma_\l$ by the 
\hbox{$(2\l+1)$-th} power of the star's radius $R$ leads to dimensionless
numbers of the type introduced by Love long ago in the Newtonian theory of tides.
In addition, one can define a third\footnote{One could also introduce magnetic-like
``shape'' Love numbers by considering other aspects of the geometry around the
  star surface.} dimensionless Love number (for any $\l$), measuring the
distortion of the {\it shape} of the surface of a star by external tidal
fields.

We have studied, both analytically and numerically, these various tidal
response coefficients, thereby generalizing a recent investigation of Flanagan
and Hinderer. The main results of our study are:
\begin{enumerate}
\item A detailed study of the strong quenching of the electric-type tidal
  coefficients $\mu_\l$ (or its dimensionless version $k_\l\sim
  G\mu_\l/R^{2\ell+1}$) as the ``compactness $c\equiv GM/(c_0^2 R)$ of the
  neutron star increases. This quenching was studied both for polytropic
  EOS (of two different types, see Fig.~\ref{fig:fig1}), 
  for the incompressible EOS (where the quenching is particularly dramatic, 
  see Fig.~\ref{fig:fig3}) and for two ``realistic'' (tabulated) EOS (see
  Fig.~\ref{fig:fig4}).

\item Part (though not all) of this quenching mechanism can be related to the
  no-hair property of black holes. The latter property ensures that some of
  the tidal response coefficients of neutron stars must vanish in the formal
  limit where $c\to c^{\rm BH}=1/2$. At face value, this suggests that the
  ``correct'' value of the $\mu_\l$ and $\sigma_\l$ tidal coefficients of
  black holes is simply zero. We, however, argued that this conclusion is
  premature, until a 5PN (5-loop) nonlinear analysis of the effective worldline
  action describing gravitationally interacting black holes is performed.

\item We gave accurate nonlinear fitting formulas for the dependence of the
  tidal coefficients $k_\ell$ and $h_\ell$ of a $\gamma=2$ $\mu$-polytrope 
  on the compactness
 (see Eqs.~\eqref{eq:fit} and \eqref{eq:hfit}). 
  We also found that two ``realistic'' EOS give rather
  close values both for the electric and magnetic tidal coefficients 
  of neutron stars. In
  particular, this suggests a possible, approximately universal analytical
  representation of the leading, quadrupolar (electric) Love number for
  neutron stars of the expected compactnesses,
  $0.12\lesssim c\lesssim 0.22$, namely
\be
\label{fit:k2}
k_2(c) \simeq 0.165-0.515 c .
\ee
Even if this simple linear fit reproduces with only a few percent accuracy the
$c$-dependence of the known realistic EOS, it might suffice to deduce from
future gravitational-wave observations an accurate value of the neutron star
compactness. Indeed, the dimensionless parameter which is crucially entering
the gravitational-wave observations is the dimensionless ratio
\be
\hat{\mu}_2(c)\equiv \dfrac{G\mu_2}{(GM/c_0^2)^5}=\dfrac{2}{3} c^{-5} k_2(c).
\ee
The strong $c$-dependence of $\hat{\mu}_2(c)$ coming from the $c^{-5}$ power
implies that even an approximate fit such as~\eqref{fit:k2} might allow one to
deduce from the measurement of $\hat{\mu}_2(c)$ a rather accurate (say to
better than $1\%$) estimate\footnote{We have in mind here, for simplicity, a
black-hole-neutron-star system where only one unknown compactness enters.} 
of $c$.

\item We surprisingly found that the magnetic-type Love numbers of neutron
  stars are negative, and quite small. We showed, by analytical arguments,
  that they can be approximately represented as $\propto B c(1-2c)$ with
  a calculable coefficient $B$ (that we computed in a few cases).

\item Following a recent investigation of the gravitational polarizability of
  black holes~\cite{DL09}, we studied the ``shape'' Love numbers $h_{\ell}$ of
  neutron stars. Again the quantity $h_\l(c)$ is found to be drastically
  quenched when $c$ increases. However, in that case $h_\l(c)$ does {\it not}
  tend to zero as $c\to c^{\rm BH}=1/2$. Rather we found that $h_\ell^{\rm NS}(c)$
  tends to the nonzero black-hole value $h^{\rm BH}_\l$~\cite{DL09} 
  as $c$ formally tends to $c^{\rm BH}=1/2$. 
\end{enumerate} 

In future work, we will come back to the other issues mentioned in the
Introduction, namely: 
\begin{enumerate}
\item the incorporation of tidal effects within the
effective one body formalism, starting from the additional term in the
effective action
\be
\Delta S = +\dfrac{1}{4}\mu_2\int ds\E_{\alpha\beta}\E^{\alpha\beta};
\ee
\item the study of the measurability of various
tidal coefficients within the signal seen by interferometric detectors of
gravitational-waves.
\end{enumerate}

After the submission of this work, a related paper by Binnington 
and Poisson~\cite{Binnington:2009bb} appeared on the archives. 
Ref.~\cite{Binnington:2009bb} develops the theory of electric and magnetic
Love numbers in a different gauge. Their results seem to be fully consistent
with ours, but are less general: (i) their treatment is limited to
e-polytropes, (ii) they did not consider the ``shape'' Love numbers, and (iii)
they do not discuss the effective action terms associated to tidal effects.

\acknowledgments

We are grateful to Luca Baiotti, Bruno Giacomazzo and 
Luciano Rezzolla for sharing with us, before publication, 
their data on inspiralling and coalescing binary neutron
stars, which prompted our interest in relativistic tidal 
properties of neutron stars.
We thank Orchidea Lecian for clarifying discussions and 
Sebastiano Bernuzzi for help with the numerical implementation of 
FPS and SLy EOS, and for first pointing out some errors 
in Ref.~\cite{Hinderer:2007mb}. We are also grateful to 
Tanja Hinderer for sending us an erratum of her paper 
before publication.

\end{document}